\documentclass{article}
\usepackage{arxiv}

\usepackage{amsmath,amsfonts,amssymb}
\usepackage{mathtools}
\usepackage[hidelinks]{hyperref}
\usepackage{algorithmic}
\usepackage{algorithm}
\usepackage{array}
\usepackage{relsize}
\usepackage{textcomp}
\usepackage{stfloats}
\usepackage{url}
\usepackage{verbatim}
\usepackage{graphicx}
\usepackage{natbib}
\usepackage{doi}
\usepackage{xcolor}
\usepackage{paralist}
\usepackage{blindtext}
\usepackage{multicol}
\usepackage{multirow}
\usepackage[subnum]{cases}
\usepackage{booktabs}
\usepackage{subcaption}

\newtheorem{theorem}{Theorem}[section]

\newtheorem{remark}[theorem]{Remark}

\newcount\Comments
\usepackage{color}
\newcommand{\kibitz}[2]{\ifnum\Comments=0\textcolor{#1}{#2}\fi}

\title{A unified dynamical modeling framework for cruise control and adaptive cruise control}

\author{
	Mingfeng Shang\\
	Civil Engineering Technology, Environmental Management and Safety\\
    Rochester Institute of Technology\\
	\texttt{mfsie@rit.edu} \\
	\And
	Shian Wang\\
	Civil, Environmental, and Architectural Engineering\\
     Electrical Engineering and Computer Science\\
	The University of Kansas\\
	\texttt{shian.wang@ku.edu}\\
}

\renewcommand{\shorttitle}{A unified dynamical modeling framework for cruise control and adaptive cruise control}

\begin{document}\sloppy
\maketitle

\begin{abstract}
Adaptive cruise control (ACC) vehicles are the first generation of automated vehicles. While fully automated vehicles are expected to benefit traffic flow, field experiments have shown that commercially available ACC vehicles may instead degrade it by reducing string stability and roadway throughput. To mitigate these effects, existing studies adjust the ACC control algorithm or introduce additional control inputs; however, few have examined the transition between the cruise control (CC) and ACC modes without modifying the ACC control algorithm itself, leaving the impacts of ACC vehicles incompletely understood. Although microscopic car-following models effectively describe the driving behavior of ACC vehicles, they capture only part of the dynamics that shape traffic flow, as they represent the ACC mode alone and omit the CC mode. To address this gap, we propose a unified dynamical model of CC and ACC that interpolates continuously between the two modes through a sigmoid weighting function, and improve traffic flow by designing the mode switching.

Based on this new model, we conduct an equilibrium and string stability analysis of the platoon, revealing the trade-off among safety, throughput, and string stability. The optimal switching threshold is designed under throughput-priority and safety-priority criteria, and compared against the threshold adopted by commercially available ACC vehicles. Numerical experiments show that, with a properly designed switching threshold, the throughput increases by up to 58.6\% and the average speed variation, a measure of speed oscillations, decreases by up to 39.7\% relative to the commercial baseline. We conclude that the excessively large switching threshold of commercially available ACC vehicles is a likely cause of their negative impact on traffic flow, and that this impact can be mitigated by properly reducing the threshold toward a safer and more string-stable regime.
\end{abstract}

\keywords{Cruise control \and Adaptive cruise control \and Car-following model \and String stability}

\section{Introduction}\label{section1}

The emergence of vehicle automation technologies on our roadways has the potential to transform the
transportation landscape. Among these technologies, adaptive cruise control (ACC) vehicles, i.e., SAE Levels 1-2~\citep{SAE2018}, are proving to be the first generation of driver-assist-enabled vehicles that are commercially available. The development of ACC dates back to the early studies on automated longitudinal control and intelligent vehicle systems~\citep{ioannou1993intelligent,shladover1995review},
and ACC is now commercially available across various vehicle manufacturers~\citep{milanes2014modeling}. As an extension of conventional cruise control (CC), which regulates the vehicle speed to a user-set desired value, ACC automatically adjusts the vehicle speed to maintain a safe following distance with the preceding vehicle~\citep{rajamani2011vehicle}. Longitudinally, car-following dynamics at the individual vehicle level may have wide-reaching implications on the overall traffic flow~\citep{talebpour2016influence}.

While fully automated vehicles are expected to benefit the overall traffic flow (e.g., increasing throughput and string stability~\citep{talebpour2016influence}), recent field experiments and simulations have shown that commercially available ACC vehicles may negatively impact traffic flow~\citep{shang2021impacts,gunter2020are,makridis2021openacc}. Specifically, they are found to be string unstable, meaning that small perturbations in the speed of a lead vehicle are amplified as they propagate upstream along a platoon of following vehicles~\citep{shang2021impacts,shang2022novel,gunter2020are}. Moreover, commercially available ACC vehicles tend to maintain larger time gaps than typical human drivers, which reduces the achievable roadway throughput~\citep{shang2021impacts}. As the market penetration of ACC vehicles continues to grow, these negative impacts on traffic stability and throughput are expected to become increasingly
pronounced~\citep{calvert2017will,shang2021impacts}.

Therefore, understanding why commercially available ACC vehicles negatively impact traffic flow is critical. Such an understanding not only explains the traffic phenomena observed in the field experiments and simulations, but also provides insights into how the negative impacts can be properly mitigated before ACC vehicles reach a high market penetration. 

To this end, several studies have proposed to adjust the ACC control methods to smooth traffic flow and maximize throughput, such as adapting the ACC driving behavior (e.g., the time gap and acceleration strength) to the surrounding traffic conditions in real time~\citep{kesting2008adaptive,spiliopoulou2018adaptive}, designing spacing policies to improve string stability and traffic flow capacity~\citep{santhanakrishnan2003spacing}, and introducing inter-vehicle communication such as cooperative ACC~\citep{ploeg2011design,milanes2014cooperative}.
Instead of changing the original control algorithms, some other studies have proposed to incorporate an additional control input to the vehicle dynamics. For example, Wang~\textit{et al.}~\citep{wang2022optimal,wang2023general} design an additional nonlinear feedback control input with which automated vehicles track a smoothed version of the disturbance from the preceding vehicle, thereby smoothing nonlinear traffic flow for a broad class of car-following models. While these approaches have been shown to be effective in simulations and controlled experiments, they require either adjusting the underlying control algorithm or imposing an additional control input on the vehicle. However, the control systems of commercially available ACC vehicles are proprietary black boxes whose control logic can only be inferred from measured driving data~\citep{gunter2019model,wang2020online}, and neither adjusting the control algorithm nor imposing an additional control input is generally feasible for researchers or road operators.

In parallel, mathematical modeling has proven effective in characterizing the driving behavior of ACC vehicles. For example, car-following models use ordinary differential equations (ODEs) to describe the motion of a following vehicle based on that of the vehicle immediately in front of it. In particular, the optimal velocity relative velocity (OVRV) model~\citep{milanes2014modeling} and the intelligent driver model (IDM)~\citep{treiber2000congested} have been calibrated with experimentally collected trajectory data and have been shown to accurately reproduce the car-following dynamics of commercially available ACC vehicles~\citep{gunter2019model,de2021calibrating,shang2022novel}.

While these models have been shown to effectively capture the car-following dynamics of ACC vehicles, they describe only the ACC mode and neglect the coexisting CC mode. In practice, an intelligent vehicle alternates between CC and ACC modes depending on the presence of a leading vehicle. Despite receiving relatively little attention in the literature, CC mode can also significantly influence traffic flow. Specifically, numerical simulations have shown that increasing the desired speed in CC mode improves highway capacity, whereas increasing the switching gap between the CC and ACC modes reduces it~\citep{shang2023capacity}. Consequently, ACC-only models provide an incomplete representation of intelligent vehicle behavior and may fail to capture the full traffic-flow dynamics arising from mode switching.

With this in mind, we consider describing the driving behavior of ACC vehicles with a unified model of CC and ACC, and improving the overall traffic flow by designing optimal switching between these two modes rather than changing the ACC control algorithm itself. Although few studies address the switching between CC and ACC directly, the switching among different control modes has been investigated within automated longitudinal control systems~\citep{ioannou1994throttle,gao2016multi}. For example, Ioannou
and Xu~\citep{ioannou1994throttle} design throttle and brake control systems for automatic vehicle following, in which a switching logic coordinates the throttle and brake controllers under a constant time headway policy. Gao~\textit{et al.}~\citep{gao2016multi} propose a switching strategy that considers the spacing, relative velocity, and acceleration to decide among the cruise, approach, and follow modes of
an ACC system. However, the switching conditions in these studies are specified as predefined rules whose thresholds are selected empirically, and how the switching impacts the resulting vehicle dynamics is not analyzed. While additional arguments are introduced to smooth the switching process~\citep{gao2016multi}, the mode switching remains a discrete jump between two controllers rather than a continuous transition, which is unrealistic for describing actual driving behavior. Furthermore, to the best of our knowledge, none of these studies capture the impacts of the switching design on traffic flow properties such as string stability, safety, and throughput.

Therefore, we propose a unified dynamical model of CC and ACC and design the optimal mode switching to improve traffic flow. The primary contributions of this study include: i) constructing a continuous and unified dynamical model that integrates the CC and ACC modes with a smooth transition to effectively describe the driving behavior of commercially available ACC vehicles; ii) conducting analytical equilibrium and string stability analyses of the unified model, and designing the mode switching to jointly integrate string stability, throughput, and safety; and iii) demonstrating with numerical experiments that ACC vehicles can improve traffic flow by properly tuning the switching design. 

The remainder of this article is outlined as follows. In Section~\ref{section3}, we develop the unified dynamical model of CC and ACC. In Section~\ref{section:equilibrium}, we analyze the equilibrium states of the unified model and their impacts on throughput. In Section~\ref{section: Stability}, we conduct a string stability analysis and derive the string stability criterion. In Section~\ref{section:optimal}, we design the optimal mode switching considering string stability, throughput, and safety. In Section~\ref{section4}, we conduct numerical experiments to evaluate the performance of the proposed switching design. Finally, we conclude in Section~\ref{section5} that the proposed switching design enables commercially available ACC vehicles to positively impact traffic
flow.

\section{A unified dynamical model of CC and ACC}\label{section3}

In realistic traffic, an intelligent vehicle alternates between CC and ACC modes depending on the surrounding traffic environment, particularly the presence of a leading vehicle. Specifically, it tends to operate in CC mode when the spacing to the preceding vehicle is sufficiently large, whereas ACC mode becomes dominant as the spacing decreases. To capture this behavior, we propose a unified car-following model that continuously transitions between CC and ACC dynamics through a sigmoid-based weighting function.

\subsection{Microscopic Vehicle Dynamics}\label{section3.1}

When a vehicle is driving alone on a roadway or sufficiently far from its preceding vehicle, it tends to operate in CC mode, whose primary objective is to maintain a user-defined desired speed. In this study, the CC dynamics are modeled using a proportional controller~\citep{gao2022optimal} given by
\begin{align}
    \dot{v}_{CC}
    =g(v)
    =
    k_P(v_d-v),
    \label{eq:cc}
\end{align}
where $v_d$ and $v$ denote the desired speed and instantaneous speed of the ego vehicle, respectively, and $k_P>0$ is the proportional control gain.

\begin{figure}
  \centering
  \includegraphics[scale=0.6, trim=7cm 10cm 7cm 7cm, clip]{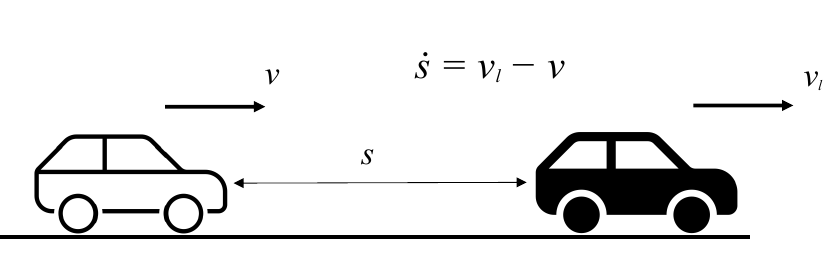}
  \vskip-5pt
  \caption{\centering An illustration of car-following behavior.}
  \label{fig:car}
\end{figure}

When the ego vehicle approaches its preceding vehicle, ACC mode becomes dominant and regulates the vehicle spacing according to a constant time headway policy. The resulting acceleration dynamics are described using the OVRV model~\citep{milanes2014modeling,gunter2019model}, given by
\begin{align}
    \dot{v}_{ACC}
    =h(s, \dot s, v)
    =
    k_1(s-\tau v-\eta)
    +
    k_2\dot s,
    \label{eq:acc}
\end{align}
where $s$ is the inter-vehicle spacing, $\dot s=v_l-v$ denotes the relative speed between the lead vehicle and the ego vehicle, $v_l$ signifies the speed of the lead vehicle, $k_1$ and $k_2$ are feedback gains associated with spacing and relative speed, respectively, $\tau$ represents the time headway, and $\eta$ denotes the jam distance. An illustration of car-following behavior is presented in Figure~\ref{fig:car}.

\subsection{A Unified Dynamical Model}\label{section3.2}

\subsubsection{Mathematical preliminaries}\label{section3.2.1}

Here we briefly introduce sigmoid functions, which have recently been utilized in modeling nonlinear car-following behavior and ACC vehicle dynamics~\citep{shang2022modeling,zhang2023interactive,chen2024sigmoid,shang2024two}. A sigmoid function is characterized by an $S$-shaped curve and possesses several important properties, including boundedness, differentiability, and monotonicity. One distinctive feature of a sigmoid function is its asymptotic behavior, constrained by a pair of horizontal asymptotes as the input variable approaches $\pm\infty$~\citep{han1995influence}. 

\begin{figure}
  \centering
  \includegraphics[scale=0.5, trim=6cm 5cm 6cm 5cm, clip]{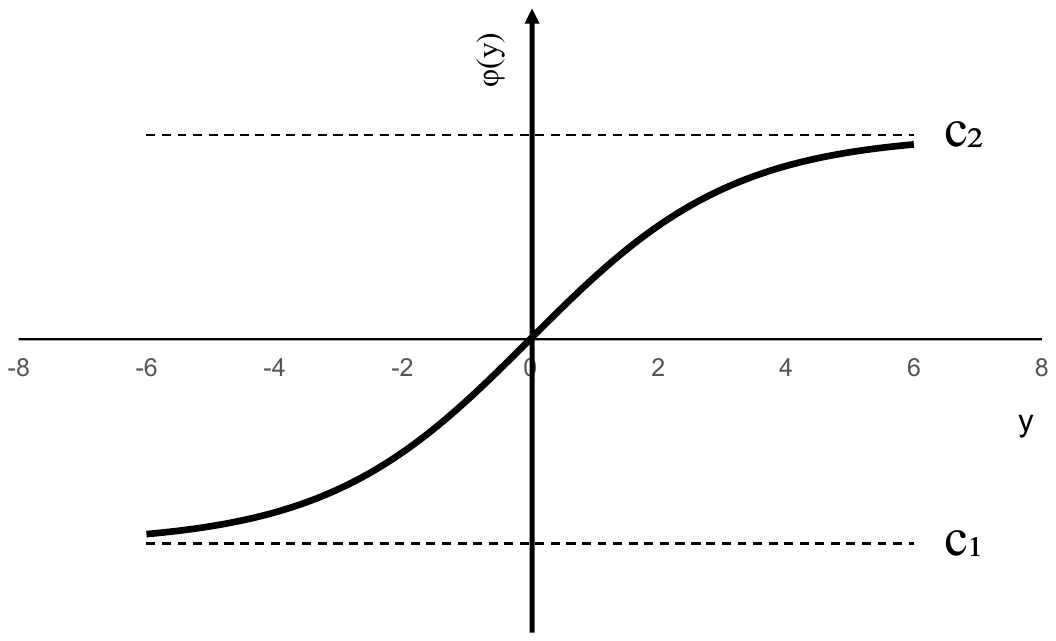}
  \vskip-20pt
  \caption{\centering An illustration of a sigmoid function $\varphi (y)$}
  \label{fig:sigmoid}
\end{figure}

Let $\varphi:\mathbb{R}\rightarrow\mathbb{R}$ denote a class of sigmoid functions satisfying $\varphi(0)=0$. We denote the corresponding lower and upper asymptotes as
\begin{subequations}\label{eq3.1}
\begin{align}
\inf \varphi(y) &= c_1,
\label{eq3.1a}\\
\sup \varphi(y) &= c_2,
\label{eq3.1b}
\end{align}
\end{subequations}
where $\inf$ and $\sup$ are the infimum and supremum of the function, respectively. A sketch of $\varphi (y)$ is shown in Figure~\ref{fig:sigmoid}.

Common examples include the hyperbolic tangent function, arctangent function, and error function. Specifically, a hyperbolic tangent sigmoid function, $\tanh(y)$, is a candidate of $\varphi$ due to its smoothness and differentiability properties. Without loss of generality, let us define
\begin{align}\label{eq:tanh}
\varphi(y)
=
\tanh(y),
\end{align}
whose lower and upper asymptotes are
\begin{align}
c_1=-1,
\qquad
c_2=1.
\end{align}

\subsubsection{Modeling}\label{section3.2.2}

To capture the continuous transition between CC and ACC modes, we define the unified vehicle dynamics as
\begin{align}\label{eq:switch}
    \dot{v}
    =
    f(s,\dot{s},v)  
    =
    \mathcal{I}(s,v)\dot{v}_{CC}
    +
    \left(1-\mathcal{I}(s,v)\right)\dot{v}_{ACC} 
    =
    \mathcal{I}(s,v)g(v)
    +
    \left(1-\mathcal{I}(s,v)\right)h(s,\dot{s},v),
\end{align}
where $g(v)$ and $h(s, \dot s, v)$ are given in~\eqref{eq:cc} and~\eqref{eq:acc}, respectively. The weighting function $\mathcal{I}(s,v)$ is defined as
\begin{align}\label{eq:Is}
\mathcal{I}(s,v)
=
\frac{1}{c_2-c_1}
\varphi
\left(
\alpha(s-s_c(v))
\right)
+
\frac{c_1}{c_1-c_2},
\end{align}
where $\alpha>0$ is a scaling factor determining the steepness of the transition, and $s_c$ denotes the critical switching spacing. It follows from straightforward algebraic manipulation that $\mathcal{I} \in (0,1)$, indicating that~\eqref{eq:switch} is indeed a convex combination of $g$ and $h$.

The above two equations can be interpreted under three representative scenarios:
\vskip3pt
\noindent\textbf{Case (i): $s\ll s_c(v)$.} In this case,
\begin{align}
\varphi\left(\alpha(s-s_c(v))\right)\rightarrow c_1,
\end{align}
which yields
\begin{align}
\mathcal{I}(s,v)\rightarrow 0.
\end{align}
Consequently,
\begin{align}
\dot{v}\rightarrow\dot{v}_{ACC}.
\end{align}
This indicates that when the inter-vehicle spacing is significantly smaller than the switching threshold, the ego vehicle primarily operates in ACC mode.

\vskip3pt
\noindent\textbf{Case (ii): $s=s_c(v)$.} In this case,
\begin{align}
\varphi\left(\alpha(s-s_c(v))\right)=0,
\end{align}
and therefore
\begin{align}
\mathcal{I}(s,v)
=
\frac{c_1}{c_1-c_2}.
\end{align}
The resulting acceleration becomes a weighted combination of CC and ACC dynamics:
\begin{align}
\dot{v}
=
\frac{c_1}{c_1-c_2}\dot{v}_{CC}
-
\frac{c_2}{c_1-c_2}\dot{v}_{ACC}.
\end{align}
This corresponds to a transition regime where the vehicle dynamics are jointly influenced by CC and ACC behaviors.

\vskip3pt
\noindent\textbf{Case (iii): $s\gg s_c(v)$.} In this case,
\begin{align}
\varphi\left(\alpha(s-s_c(v))\right)\rightarrow c_2,
\end{align}
which yields
\begin{align}
\mathcal{I}(s,v)\rightarrow 1.
\end{align}
Consequently,
\begin{align}
\dot{v}\rightarrow\dot{v}_{CC}.
\end{align}
This indicates that when the inter-vehicle spacing is significantly larger than the switching threshold, the vehicle predominantly operates in CC mode.

Assume that the critical switching spacing $s_c$ is linearly related to the operating speed $v$; we define the switching threshold as
\begin{align}\label{eq:sc}
s_c(v)
=
s_0+\tau_c v,
\end{align}
where $s_0$ denotes the minimum switching spacing (the switching threshold at standstill) and $\tau_c$ is the switching-threshold time-headway parameter. Smaller values of $\tau_c$ promote CC operation, whereas larger values favor ACC operation. The linear form in~\eqref{eq:sc} captures the dominant speed dependence of the switching threshold while preserving analytical tractability. In addition, $\tau_c$ admits a direct physical interpretation as a time headway, facilitating practical calibration.

Recall the tangent function, substituting the hyperbolic tangent~\eqref{eq:tanh} into~\eqref{eq:Is} leads to the explicit weighting function
\begin{align}\label{eq:Is2}
\mathcal{I}(s,v)
=
\frac12\left[1+\tanh\left(\alpha(s-s_c(v))\right)\right].
\end{align}
Using $\tfrac{d}{dy}\tanh(y)=\operatorname{sech}^2(y)$, the derivative of the weighting function with respect to spacing is
\begin{align}\label{eq:Isprime}
\mathcal{I}_s(s,v)
=
\frac{\partial \mathcal{I}(s,v)}{\partial s}
=
\frac{\alpha}{2}\operatorname{sech}^2\left(\alpha(s-s_c(v))\right).
\end{align}
Furthermore, using $\operatorname{sech}^2(y)=1-\tanh^2(y)$ together with~\eqref{eq:Is2},~\eqref{eq:Isprime} can be written in the compact form
\begin{align}\label{eq:Iprime2}
\mathcal{I}_s(s,v)
=
2\alpha\,\mathcal{I}(s,v)\left(1-\mathcal{I}(s,v)\right).
\end{align}

\begin{remark}
Note that $\tau_c$ is distinct from the ACC desired time headway $\tau$. Specifically, $\tau$ represents the time headway within ACC dynamics, whereas $\tau_c$ determines how conservatively the vehicle transitions from ACC-dominant operation to CC-dominant operation.   
\end{remark}

\subsubsection{Safety-constrained switching}\label{section:safety}

Since a small $\tau_c$ may result in an insufficient switching gap $s_c$ and compromise safety by allowing CC-dominant behavior at small spacings, $\tau_c$ should satisfy appropriate safety constraints. To ensure collision-free operation, a braking-based safe-spacing condition is introduced as follows
\begin{align}\label{eq:safe}
s_{\mathrm{safe}}(v,v_l)
=
s_{\min}
+
vt_r
+
\frac{v^2-v_l^2}{2\beta},
\end{align}
where $s_{\min}$ is the minimum standstill safe spacing, $t_r$ denotes the reaction (or actuation) delay, $\beta$ is the comfortable braking deceleration, assumed to be identical for the lead and following vehicles, and $v_l$ is the lead-vehicle speed. Equation~\eqref{eq:safe} is obtained by simplifying the responsibility-sensitive safety (RSS) longitudinal safe distance~\citep{shalevshwartz2017rss} under the assumptions of equal braking deceleration and constant lead-vehicle speed during the reaction interval, with an additional standstill spacing $s_{\min}$. Physically,~\eqref{eq:safe} requires the ego vehicle to maintain sufficient spacing to avoid a collision if both vehicles brake at rate $\beta$ after a reaction delay of $t_r$. Here, $\beta$ represents a nominal comfortable braking deceleration. Using the same value for both vehicles simplifies the analysis and allows the proposed switching framework to be isolated from differences in vehicle braking capabilities.

To ensure CC-dominant behavior remains within the safe region, the switching threshold must satisfy
\begin{align}\label{eq:safe2}
s_c(v)
\ge
s_{\mathrm{safe}}(v,v_l).
\end{align}

Consider steady-state following, in which the ego vehicle speed equals the leader's speed ($v=v_l$); then $s_{\mathrm{safe}}=s_{\min}+vt_r$. Substituting~\eqref{eq:sc} into~\eqref{eq:safe2} and requiring the safety condition to hold at every following speed $v\ge0$ gives
\begin{align}\label{eq:safe3}
s_0+\tau_c v
\ge
s_{\min}+t_r v,
\quad \forall\, v\ge 0.
\end{align}

Being affine in $v$, this holds if and only if $s_0\ge s_{\min}$ (the constant term, at $v=0$) and $\tau_c\ge t_r$ (the slope). Choosing the tightest switching offset $s_0=s_{\min}$, for which CC engagement begins exactly at the safe spacing when stationary, the safety requirement reduces to the speed-independent bound
\begin{align}\label{eq:tauc_tr}
\tau_c \ge t_r.
\end{align}

\section{Equilibrium state analysis}\label{section:equilibrium}

In this section, we discuss the equilibrium state of traffic flow to understand traffic efficiency. At equilibrium, all vehicles travel at a constant speed ($v^*$) with identical spacing ($s^*$), meaning
\begin{align}
\dot{v}_i^*=0,
\qquad
\dot{s}_i^*=0,
\end{align}
for all vehicles.

Substituting the unified dynamics of~\eqref{eq:switch} into the equilibrium condition yields
\begin{align}\label{eq:eq_condition}
0
=
\mathcal{I}(s^*,v^*)
k_P(v_d-v^*)  
+ 
\left(
1-\mathcal{I}(s^*,v^*)
\right)
k_1(s^*-\tau v^*-\eta),
\end{align}
where $s^*$ and $v^*$ denote the equilibrium spacing and equilibrium speed, respectively.

Define the equilibrium time headway as
\begin{align}
\theta^*
=
\frac{s^*+l}{v^*},
\end{align}
where $l$ is the vehicle length. 

Substituting $s^*=\theta^*v^*-l$ into~\eqref{eq:eq_condition} leads to
\begin{align}
0
=
\mathcal{I}^*
k_P(v_d-v^*)
+
(1-\mathcal{I}^*)
k_1
\left(
v^*(\theta^*-\tau)-l - \eta
\right),
\end{align}
where
\begin{align}
\mathcal{I}^*
=
\mathcal{I}(s^*,v^*).
\end{align}

Solving for $\theta^*$ gives
\begin{align}\label{eq:theta}
\theta^*
=
\tau
+
\frac{\eta+l}{v^*}
-
\frac{k_P(v_d-v^*)}{k_1v^*}
\cdot
\frac{\mathcal{I}^*}{1-\mathcal{I}^*}.
\end{align}

The equilibrium throughput is therefore given by
\begin{align}\label{eq:q}
q^*
=
\frac{v^*}{s^*+l}
=
\frac{1}{\theta^*}.
\end{align}

\begin{remark}\label{remark:vd}
We assume $v_d>v^*$, where $v_d$ represents the desired free-flow speed in the absence of a preceding vehicle. Since the equilibrium speed $v^*$ is determined by car-following interactions and is constrained by the presence of a leader vehicle, it is generally lower than the unconstrained desired speed $v_d$.
\end{remark}

\begin{remark}\label{remark:throughput}
For an equilibrium operating speed $v^*$, $\theta^*$ in~\eqref{eq:theta} monotonically decreases with respect to the equilibrium weighting factor $\mathcal{I}^*$ under the condition $v_d>v^*$. Consequently, the equilibrium throughput $q^*=1/\theta^*$ monotonically increases with $\mathcal{I}^*$ for $\theta^*>0$. Therefore, maximizing the equilibrium throughput is equivalent to maximizing the admissible value of $\mathcal{I}^*$. This relationship can be verified by differentiating~\eqref{eq:theta} which yields
$d\theta^*/d\mathcal{I}^*=-(k_P/k_1)(v_d-v^*)/[v^*(1-\mathcal{I}^*)^2]<0$
for $v_d>v^*$. Since $dq^*/d\mathcal{I}^*=-(\theta^*)^{-2}\,d\theta^*/d\mathcal{I}^*>0$, the throughput increases monotonically with $\mathcal{I}^*$ when $\theta^*>0$. 
\end{remark}

\begin{remark}\label{remark:safety}
It follows from~\eqref{eq:safe3} that the braking-based safety constraint requires
\begin{align}
s_c(v^*)
\ge
s_{\rm safe}(v^*)
=
s_{\min}+v^*t_r.
\end{align}

Furthermore,
\begin{align}
\frac{\partial \mathcal{I}(s,v)}
{\partial s_c}
=
-
\frac{\alpha}{c_2-c_1}
\varphi'
\!\left(
\alpha(s-s_c(v))
\right)
<0,
\end{align}
indicating that the switching function monotonically decreases with the switching threshold. Consequently, the safety constraint implies
\begin{align}
\mathcal{I}^*
\le
\frac{1}{c_2-c_1}
\varphi
\!\left(
\alpha
\left(
s^*
-
s_{\rm safe}(v^*)
\right)
\right)
+
\frac{c_1}{c_1-c_2}
:=\mathcal{I}_{\max}^*.
\end{align}

Since
\begin{align}
\varphi(y)<c_2,
\quad
\forall y\in\mathbb{R},
\end{align}
it follows immediately that
\begin{align}
\mathcal{I}_{\max}^*<1.
\end{align}

Hence,
\begin{align}
0<\mathcal{I}^*\le \mathcal{I}_{\max}^*<1.
\end{align}

Therefore, the braking-based safety constraint imposes an explicit upper bound $\mathcal{I}_{\max}^*$ on $\mathcal{I}^*$, and hence on the equilibrium throughput.
\end{remark}

Combining the monotonicity of throughput in $\mathcal{I}^*$ (Remark~\ref{remark:throughput}) with the safety cap $\mathcal{I}^*\le \mathcal{I}_{\max}^*$ (Remark~\ref{remark:safety}), three representative operating regimes emerge:
\vskip3pt
\noindent\textbf{Pure ACC regime ($\mathcal{I}^* \rightarrow 0$).} In this regime,
\begin{align}
\theta^*
=
\tau+\frac{\eta+l}{v^*},
\end{align}
where the equilibrium spacing is primarily determined by the constant time-headway policy, resulting in limited throughput. 

\vskip3pt
\noindent\textbf{CC-dominant regime ($\mathcal{I}^* \rightarrow \mathcal{I}_{\max}^*$).} As $\mathcal{I}^*$ increases, CC mode becomes increasingly dominant and the equilibrium time headway decreases, leading to higher traffic throughput. However, $\mathcal{I}^*$ cannot exceed $\mathcal{I}_{\max}^*$ due to the braking-based safety constraint. Therefore, the maximum achievable throughput is limited by the safety requirement.

\vskip3pt
\noindent\textbf{Transition regime ($0<\mathcal{I}^*<\mathcal{I}_{\max}^*$).} Within this regime, the proposed unified model continuously interpolates between ACC-dominant and CC-dominant behaviors. Consequently, the transition regime establishes a continuous tradeoff between traffic throughput and safety.

\section{String stability analysis}\label{section: Stability}

In this section, we analyze the string stability of the unified model. Recall from Section~\ref{section:equilibrium} that at equilibrium all vehicles travel at the same constant speed $v^*$ with identical spacing $s^*$, so that $\dot v_i=\dot s_i=0$ and $f(s^*,0,v^*)=0$.

To characterize the local dynamics around the equilibrium, the unified model is linearized as
\begin{align}
\ddot{x}_i
\approx
\left.\frac{\partial f}{\partial s}\right|_{eq}(s_i-s^*)
+
\left.\frac{\partial f}{\partial \dot{s}}\right|_{eq}\dot{s}_i
+
\left.\frac{\partial f}{\partial v}\right|_{eq}(v_i-v^*).
\end{align}

Define
\begin{align}
k_s
&=
\left.\frac{\partial f}{\partial s}\right|_{eq},\\
k_d
&=
\left.\frac{\partial f}{\partial \dot{s}}\right|_{eq},\\
k_v
&=
-
\left.\frac{\partial f}{\partial v}\right|_{eq}.
\end{align}

Let $l$ denote the vehicle length. The inter-vehicle spacing is defined as
\begin{align}
s_i(t)
=
x_{i-1}(t)-x_i(t)-l.
\end{align}

Assuming the platoon is initialized at the equilibrium spacing, this initial gap equals $s^*$,
\begin{align}
s^*
=
x_{i-1}(0)-x_i(0)-l.
\end{align}

Define the perturbation state as
\begin{align}
\tilde{x}_i(t)
=
x_i(t)-x_i(0)-tv^*.
\end{align}

Using the above definitions,
\begin{align}
s_i(t)-s^*
&=
\tilde{x}_{i-1}(t)-\tilde{x}_i(t),\\
\dot{s}_i(t)
&=
\dot{\tilde{x}}_{i-1}(t)-\dot{\tilde{x}}_i(t),\\
v_i(t)-v^*
&=
\dot{\tilde{x}}_i(t).
\end{align}

Substituting these relationships into the linearized
dynamics yields
\begin{align}
\ddot{\tilde{x}}_i
=
k_s(\tilde{x}_{i-1}-\tilde{x}_i)
+
k_d(\dot{\tilde{x}}_{i-1}-\dot{\tilde{x}}_i)
-
k_v\dot{\tilde{x}}_i.
\end{align}

Assuming zero initial perturbations and letting $z$ denote the Laplace variable, the Laplace transform gives
\begin{align}
T(z)
:=
\frac{\tilde X_i(z)}
{\tilde X_{i-1}(z)}
=
\frac{k_s+k_dz}
{z^2+(k_d+k_v)z+k_s}.
\end{align}

\begin{remark}
Following the classical definition of local string stability~\citep{wilson2011car}, the system is locally string stable if and only if
\begin{align}
|T(j\omega)|
\le
1,
\quad
\forall\omega\in\mathbb{R}^+.
\end{align}
\end{remark}

Substituting the transfer function into the above condition and comparing the numerator and denominator yields
\begin{align}\label{eq:ss_criterion}
k_s
\le
k_dk_v+\frac12k_v^2,
\end{align}
which serves as the local string stability criterion of the proposed unified model.

We now evaluate the gains $k_s$, $k_d$, and $k_v$ by differentiating the unified model~\eqref{eq:switch}. The partial derivatives of the weighting function are denoted by
\begin{align}
\mathcal{I}_s(s,v)
=
\frac{\partial \mathcal{I}(s,v)}{\partial s},
\qquad
\mathcal{I}_v(s,v)
=
\frac{\partial \mathcal{I}(s,v)}{\partial v}.
\end{align}

It follows from~\eqref{eq:Is} and~\eqref{eq:sc} that the partial derivatives of $\mathcal{I}(s,v)$ satisfy
\begin{align}\label{eq:Iv_Is}
\mathcal{I}_v(s,v)
=
-s_c'(v)\mathcal{I}_s(s,v)
=
-\tau_c\mathcal{I}_s(s,v).
\end{align}

Based on $dg/dv=-k_P$ from~\eqref{eq:cc} and the partial derivatives $\partial h/\partial s=k_1$, $\partial h/\partial\dot{s}=k_2$, and $\partial h/\partial v=-k_1\tau$ from~\eqref{eq:acc}, along with~\eqref{eq:Iv_Is}, the partial derivatives of $f$ can be obtained as
\begin{align}
\frac{\partial f}{\partial s}
&=
\mathcal{I}_s(s,v)\big(g(v)-h(s,\dot{s},v)\big)
+
k_1\big(1-\mathcal{I}(s,v)\big),\\
\frac{\partial f}{\partial \dot{s}}
&=
k_2\big(1-\mathcal{I}(s,v)\big),\\
\frac{\partial f}{\partial v}
&=
-\tau_c\mathcal{I}_s(s,v)\big(g(v)-h(s,\dot{s},v)\big)
-
k_P\mathcal{I}(s,v)
-
k_1\tau\big(1-\mathcal{I}(s,v)\big).
\end{align}

Evaluating at the equilibrium $(s^*,0,v^*)$ and using $k_s=\left.\partial f/\partial s\right|_{eq}$,
$k_d=\left.\partial f/\partial\dot{s}\right|_{eq}$, and $k_v=-\left.\partial f/\partial v\right|_{eq}$ gives
\begin{align}
k_s
&=
\mathcal{I}_s^*\big(g(v^*)-h(s^*,0,v^*)\big)
+
k_1(1-\mathcal{I}^*),\\
k_d
&=
k_2(1-\mathcal{I}^*),\\
k_v
&=
\tau_c\mathcal{I}_s^*\big(g(v^*)-h(s^*,0,v^*)\big)
+
k_P\mathcal{I}^*
+
k_1\tau(1-\mathcal{I}^*),
\end{align}
where $\mathcal{I}^*=\mathcal{I}(s^*,v^*)$ and
$\mathcal{I}_s^*=\mathcal{I}_s(s^*,v^*)$.

\begin{remark}
The proposed unified model shows three representative string-stability regimes:
\vskip3pt
\noindent\textbf{ACC-dominant regime ($s\ll s_c(v)$).} In this regime,
\begin{align}
\mathcal{I}(s,v)\rightarrow 0,
\qquad
\mathcal{I}_s(s,v)\rightarrow 0.
\end{align}
Since $\mathcal{I}_v(s,v)=-\tau_c \mathcal{I}_s(s,v)$, we also have
\begin{align}
\mathcal{I}_v(s,v)\rightarrow 0.
\end{align}
Consequently,
\begin{align}
k_v\rightarrow k_1\tau,
\qquad
k_d\rightarrow k_2,
\qquad
k_s\rightarrow k_1.
\end{align}
Substituting these expressions into~\eqref{eq:ss_criterion} yields
\begin{align}
1
\le
k_2\tau+\frac12k_1\tau^2,
\end{align}
which recovers the classical string stability condition for
constant-time-headway ACC models.

\vskip3pt
\noindent\textbf{CC-dominant regime ($s\gg s_c(v)$).} In this regime,
\begin{align}
\mathcal{I}(s,v)\rightarrow 1,
\qquad
\mathcal{I}_s(s,v)\rightarrow 0,
\qquad
\mathcal{I}_v(s,v)\rightarrow 0.
\end{align}
Therefore,
\begin{align}
k_v\rightarrow k_P,
\qquad
k_d\rightarrow 0,
\qquad
k_s\rightarrow 0.
\end{align}
The vehicle dynamics become independent of the preceding vehicle and behave as a desired-speed tracking controller.

\vskip3pt
\noindent\textbf{Transition regime ($s\approx s_c(v)$).} Since $0<\mathcal{I}(s,v)<1$, $\mathcal{I}_s(s,v)=2\alpha\mathcal{I}(s,v)(1-\mathcal{I}(s,v))>0$ by~\eqref{eq:Iprime2}. As $\mathcal{I}_s$ no longer vanishes, the spacing gain $k_s$ and the speed-feedback gain $k_v$ each acquire an extra switching-induced term: $\mathcal{I}_s(s,v)\big(g(v)-h(s,\dot{s},v)\big)$ in the former and, since $\mathcal{I}_v=-\tau_c\mathcal{I}_s$, $\tau_c\mathcal{I}_s(s,v)\big(g(v)-h(s,\dot{s},v)\big)$ in the latter. These terms do not exist in the ACC- and CC-dominant regimes; they peak near the switching boundary, where $\mathcal{I}_s$ is largest, with magnitude determined by the sigmoid steepness $\alpha$ and the switching time-headway $\tau_c$.
\end{remark}

\section{Optimal switching design}\label{section:optimal}

The design objective is to maximize equilibrium throughput while satisfying string stability and safety constraints for the vehicle platoon. As established in Sections~\ref{section:equilibrium} and~\ref{section: Stability}, these properties are all determined by the equilibrium weighting factor $\mathcal{I}^*=\mathcal{I}(s^*,v^*)$. The following subsections analyze these three design criteria.

\subsection{Throughput-Oriented Objective}\label{section:5.1}

For the design, it is convenient to rewrite the equilibrium time headway~\eqref{eq:theta} as
\begin{align}\label{eq:theta_HM}
\theta^*
=
H
-
M\frac{\mathcal{I}^*}{1-\mathcal{I}^*},
~
H
=
\tau+\frac{\eta+l}{v^*},
~
M
=
\frac{k_P(v_d-v^*)}{k_1v^*}.
\end{align}
Since $M>0$ for $v_d>v^*$, Remark~\ref{remark:throughput} indicates that the throughput $q^*=1/\theta^*$ increases monotonically with $\mathcal{I}^*$; hence maximizing throughput is equivalent to maximizing $\mathcal{I}^*$.

\subsection{Safety Constraints}\label{section:5.2}

The braking-based safety constraint of Remark~\ref{remark:safety} requires the switching spacing to satisfy $s_c(v^*)\ge s_{\rm safe}(v^*)=s_{\min}+v^*t_r$. Since the weighting function monotonically decreases with the switching threshold, this yields an upper bound on the equilibrium weighting factor,
\begin{align}\label{eq:Imax}
0<\mathcal{I}^*\le\mathcal{I}_{\max}^*<1,
\end{align}
with $\mathcal{I}_{\max}^*$ given in Remark~\ref{remark:safety}. 

\subsection{String-Stability Constraints}\label{section:5.3}

From Section~\ref{section: Stability}, the local string stability condition is
\begin{align}
k_s
\le
k_dk_v+\frac12k_v^2.
\end{align}
Equivalently, we define the string-stability margin as
\begin{align}
\Sigma(\mathcal{I}^*)
=
k_dk_v+\frac12k_v^2-k_s.
\end{align}
The system is locally string stable if
\begin{align}
\Sigma(\mathcal{I}^*)\ge0.
\end{align}

Using the coefficients derived in Section~\ref{section: Stability}, let
\begin{align}
\Delta^*
=
g(v^*)-h(s^*,0,v^*).
\end{align}
It follows that
\begin{align}
k_s
&=
\mathcal{I}_s^*\Delta^*
+
k_1(1-\mathcal{I}^*),\\
k_d
&=
k_2(1-\mathcal{I}^*),\\
k_v
&=
\tau_c \mathcal{I}_s^*\Delta^*
+
k_P\mathcal{I}^*
+
k_1\tau(1-\mathcal{I}^*).
\end{align}

At equilibrium, $\mathcal{I}^*g(v^*)+(1-\mathcal{I}^*)h(s^*,0,v^*)=0$, hence $h(s^*,0,v^*)=-\tfrac{\mathcal{I}^*}{1-\mathcal{I}^*}g(v^*)$ and
\begin{align}\label{eq:Delta_star}
\Delta^*
=
g(v^*)-h(s^*,0,v^*)
=
\frac{g(v^*)}{1-\mathcal{I}^*}
=
\frac{k_P(v_d-v^*)}{1-\mathcal{I}^*}.
\end{align}
It follows from~\eqref{eq:Iprime2} that $\mathcal{I}_s^*=2\alpha \mathcal{I}^*(1-\mathcal{I}^*)$. Substituting this and $\Delta^*$ into $k_s,k_d,k_v$ cancels the factor $(1-\mathcal{I}^*)$ and leaves all three gains affine in $\mathcal{I}^*$:
\begin{align}
k_s &= k_1+(A-k_1)\,\mathcal{I}^*,  \label{eq:affine_gains_ks}\\
k_d &= k_2-k_2\,\mathcal{I}^*, \label{eq:affine_gains_kd}\\
k_v &= k_1\tau+(\tau_c A+k_P-k_1\tau)\,\mathcal{I}^*,  \label{eq:affine_gains_kv}
\end{align}
with $A:=2\alpha k_P(v_d-v^*)>0$. The stability margin $\Sigma(\mathcal{I}^*)=k_dk_v+\tfrac12k_v^2-k_s$ is therefore a quadratic in $\mathcal{I}^*$, with endpoint values $\Sigma(0)=k_1\!\left(k_2\tau+\tfrac12k_1\tau^2-1\right)$ and $\Sigma(1)=\tfrac12\left(k_P+\tau_c A\right)^2-A$. It is easy to verify that $\Sigma(1)$ is positive for a sufficiently small sigmoid steepness $\alpha$, since $A=2\alpha k_P(v_d-v^*) \to 0$ and $\Sigma(1)\to\frac{1}{2}k_P^2>0$ as $\alpha \to 0$.

As indicated for the string stability of commercially available ACC vehicles~\citep{shang2021impacts,gunter2019modeling}, the calibrated ACC gains satisfy $k_2\tau+\tfrac12k_1\tau^2<1$, so $\Sigma(0)<0$ and the commercially available ACC is string unstable. Therefore, a quadratic with $\Sigma(0)<0<\Sigma(1)$ has a unique root $\mathcal{I}_{\rm stab}^*\in(0,1)$, hence the platoon is string stable if and only if
\begin{align}\label{eq:Istab}
\mathcal{I}^*\ge \mathcal{I}_{\rm stab}^*.
\end{align}
String stability therefore sets a lower bound on the weighting factor $\mathcal{I}^*$.

\subsection{Feasible Region and Optimal Design}\label{section:5.4}

\subsubsection{Feasible region of the weighting factor $\mathcal{I}^*$}
Combining the string-stability lower bound~\eqref{eq:Istab} with the safety upper bound~\eqref{eq:Imax}, a feasible equilibrium weighting factor must satisfy
\begin{align}\label{eq:feasible_I}
\mathcal{I}_{\rm stab}^*\le \mathcal{I}^*\le \mathcal{I}_{\max}^*.
\end{align}
Hence, a feasible design exists if and only if
\begin{align}
\mathcal{I}_{\rm stab}^*\le \mathcal{I}_{\max}^*.
\end{align}
In other words, every point in the feasible window $[\mathcal{I}^*_{\rm stab},\mathcal{I}^*_{\max}]$ is safe and string stable. Within it, both the throughput $q^*$ and the string-stability margin increase with $\mathcal{I}^*$, whereas the safety margin decreases.

\subsubsection{Optimal design of switching parameters}

The design point depends on the selected design objective. Since both the equilibrium throughput and the string-stability margin increase monotonically with $\mathcal{I}^*$, the throughput-priority and stability-priority designs coincide at the upper bound, $\mathcal{I}^*=\mathcal{I}^*_{\max}$, where the safety constraint becomes active. In contrast, the safety-priority design selects the lower bound, $\mathcal{I}^*=\mathcal{I}^*_{\rm stab}$, which maximizes the safety margin while placing the platoon at the string-stability boundary.

For the throughput-priority and stability-priority optimal design, 
\begin{align}
\mathcal{I}^{*,opt}=\mathcal{I}_{\max}^*,
\end{align}
which yields $s_c^{opt}(v^*)=s_{\rm safe}(v^*)=s_{\min}+v^*t_r$; under the design convention $s_0=s_{\min}$, the switching threshold attains its minimum admissible value,
\begin{align}
\tau_c^{opt}=t_r.
\end{align}
Decreasing $\mathcal{I}^*$ toward $\mathcal{I}^*_{\rm stab}$, equivalently increasing $\tau_c$, sacrifices throughput and stability margin in favor of a safety margin. Note that this optimum is guaranteed only when the feasibility condition~\eqref{eq:feasible_I} holds; if it fails, no safe and string-stable design exists.

\begin{remark}[Stability boundary $\mathcal{I}^*_{\rm stab}$ and switching threshold $\tau_{c,\rm stab}$]\label{remark:tauc_stab}
The boundary $\mathcal{I}^*_{\rm stab}$ and the associated critical threshold $\tau_{c,\rm stab}$ admit an analytical characterization.

Substituting~\eqref{eq:affine_gains_ks}--\eqref{eq:affine_gains_kv} into the string-stability criterion~\eqref{eq:ss_criterion}, the stability margin $\Sigma(\mathcal{I}^*)=k_dk_v+\tfrac12k_v^2-k_s$ becomes the quadratic
\begin{align}\label{eq:Sigma_abc}
\Sigma(\mathcal{I}^*)=a\mathcal{I}^{*2}+b\mathcal{I}^*+c,
\end{align}
with coefficients
\begin{align}
a&=(k_P-k_1\tau+\tau_c A)\!\left[\tfrac12(k_P-k_1\tau+\tau_c A)-k_2\right],\\
b&=k_2(k_P-2k_1\tau+\tau_c A)+k_1\tau(k_P-k_1\tau+\tau_c A)-A+k_1,\\
c&=k_1\!\left(k_2\tau+\tfrac12k_1\tau^2-1\right).
\end{align}
Therefore, 
\begin{align}
\mathcal{I}^*_{\rm stab}=\frac{-b \pm \sqrt{b^2-4ac}}{2a},
\end{align}

As mentioned in Section~\ref{section:5.3}, only the root $\mathcal{I}^*_{\rm stab} \in (0,1)$ is admissible. Because $\Sigma(0)=c<0$ and $\Sigma(1)=\tfrac12\left(k_P+\tau_c A\right)^2-A>0$ for sufficiently small $\alpha$, the quadratic $\Sigma(\mathcal{I}^*)$ changes sign exactly once on $(0,1)$. Thus, this single crossing corresponds to the admissible root, while the other root lies outside $(0,1)$ and is discarded.

Writing the equilibrium weighting function explicitly, $\mathcal{I}^*=\tfrac12\big[1+\tanh\!\big(\alpha(s^*-s_c(v^*))\big)\big]$ with $s_c(v^*)=s_0+\tau_c v^*$, and inverting for $\tau_c$ at $\mathcal{I}^*=\mathcal{I}^*_{\rm stab}$ gives the critical switching threshold
\begin{align}\label{eq:tauc_stab}
\tau_{c,\rm stab}=\frac{1}{v^*}\!\left[\,s^*(\mathcal{I}^*_{\rm stab})-s_0-\frac{1}{\alpha}\tanh^{-1}\!\big(2\mathcal{I}^*_{\rm stab}-1\big)\right],
\end{align}
where $s^*(\mathcal{I}^*)=\theta^*(\mathcal{I}^*)v^*-l$ is fixed by the equilibrium headway~\eqref{eq:theta}, so that $s^*$ is determined once $\mathcal{I}^*$ is specified. String stability holds if and only if $\tau_c\le\tau_{c,\rm stab}$. Since $a$ and $b$ depend on $\tau_c$, while $\tau_{c,\rm stab}$ is also determined by $\mathcal{I}^*_{\rm stab}$, the problem reduces to a scalar fixed-point iteration. Starting from $\tau_c=0$, we iteratively compute $\mathcal{I}^*_{\rm stab}$ and update $\tau_c$ according to~\eqref{eq:tauc_stab} until convergence.
\end{remark}

\begin{remark}[Effect of the sigmoid steepness $\alpha$]\label{remark:alpha}
The steepness $\alpha$ influences the stability margin through $A=2\alpha k_P(v_d-v^*)>0$, which appears in the coefficients $a$ and $b$ of the quadratic~\eqref{eq:Sigma_abc}. Differentiating~\eqref{eq:Sigma_abc} at a fixed weighting factor $\mathcal{I}^*$ yields the compact form
\begin{align}\label{eq:dSigma_dalpha}
\frac{\partial\Sigma}{\partial\alpha}
=2k_P(v_d-v^*)\,\mathcal{I}^*\big[\tau_c(k_d+k_v)-1\big].
\end{align}
When $\tau_c(k_d+k_v)<1$, the derivative in~\eqref{eq:dSigma_dalpha} is negative, so a larger $\alpha$ reduces string stability. This effect is not monotonic, as it depends on $\tau_c(k_d+k_v)$. For a small $\alpha$ (gradual switching), $A$ stays small, and therefore string stability barely depends on $\alpha$.
\end{remark}

\section{Numerical results}\label{section4}

In this section, we conducted numerical analysis to demonstrate the effectiveness of the unified model. 

\subsection{Simulation Setup}\label{section:sim_setup}

As mentioned in Section~\ref{section3.2}, we adopt the hyperbolic tangent as the sigmoid function, $\varphi(y)=\tanh(y)$, for which $c_1=-1$ and $c_2=1$, in order that the weighting function adopts to the form in~\eqref{eq:Is2} with the speed-dependent threshold $s_c(v)=s_0+\tau_c v$. 

The ACC parameters $[k_1,k_2,\tau,\eta]$ of the OVRV model are calibrated against field-measured trajectory data~\citep{gunter2020are}. The minimum switching spacing equals the minimum standstill safe spacing ($s_0=s_\text{min}=2$~m), adopted from~\citep{treiber2000congested}. The comfortable deceleration $b=3.4~\mathrm{m/s^2}$ is taken from the geometric design guidance~\citep{aashto2018}. The delay $t_r=0.5$~s represents the typical sensing delay of ACC systems~\citep{rajamani2011vehicle,zhang2025anticipatory}, and the actuation lag and input delay of the vehicle mechanical systems~\citep{elbaklish2025driving}. The CC gain $k_P=0.40~\mathrm{s^{-1}}$ is adopted from~\citep{gao2022optimal}, and the desired cruising speed $v_d=25.0~\mathrm{m/s}$ aligns with \cite{shang2021impacts}. The sigmoid steepness is chosen as $\alpha=0.03~\mathrm{m^{-1}}$ so that the CC-ACC transition occurs smoothly. All parameter values are summarized in Table~\ref{tab:params}.

\begin{table}
\centering
\caption{Model and simulation parameters.}
\label{tab:params}
\begin{tabular}{llc}
\toprule
\textbf{Symbol} & \textbf{Description} & \textbf{Value} \\
\midrule
$k_1$      & ACC spacing gain          & $0.053~\mathrm{s^{-2}}$\\
$k_2$      & ACC relative-speed gain   & $0.293~\mathrm{s^{-1}}$\\
$\tau$     & ACC desired time headway  & $0.937~\mathrm{s}$\\
$\eta$     & ACC stopping distance     & $10.8~\mathrm{m}$\\
$k_P$      & CC proportional gain      & $0.40~\mathrm{s^{-1}}$\\
$\alpha$   & Sigmoid steepness         & $0.03~\mathrm{m^{-1}}$ \\
$s_0$      & Minimum switching spacing  & $2.0~\mathrm{m}$\\
$t_r$      & Reaction/actuation delay  & $0.5~\mathrm{s}$\\
$b$        & Comfortable deceleration  & $3.4~\mathrm{m/s^2}$\\
$l$        & Vehicle length            & $5.0~\mathrm{m}$ \\
$v_d$      & Desired cruising speed    & $25.0~\mathrm{m/s}$\\
$\Delta t$ & Integration step          & $0.1~\mathrm{s}$ \\
$N$        & Platoon size              & $10$ \\
\bottomrule
\end{tabular}
\end{table}

\subsection{Relationship Between $\theta^*$ and $\mathcal{I}^*$}\label{section:sim_theta}

This subsection numerically illustrates how the weighting factor $\mathcal{I}^*$ influences the equilibrium throughput. Using the parameters in Table~\ref{tab:params}, we sweep the weighting factor $\mathcal{I}^*$ over $(0,1)$ and, for each value, evaluate the equilibrium time headway $\theta^*$ from~\eqref{eq:theta} together with the braking-based safety bound $\mathcal{I}^*_{\max}$ from Remark~\ref{remark:safety}. The objective is to quantify the throughput changes from the ACC-dominant regime toward the CC-dominant regime, and to identify the bound due to the safety limit.

\begin{figure}
  \centering
  \includegraphics[width=0.5\columnwidth]{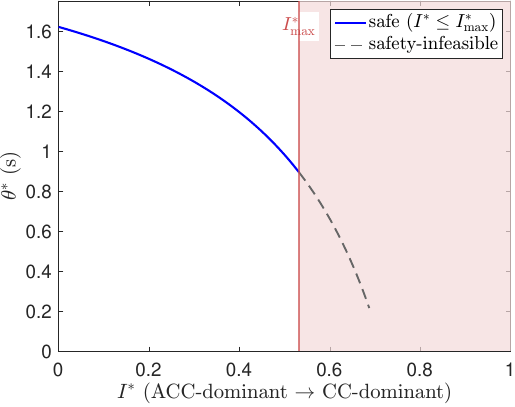}
  \caption{Equilibrium time headway $\theta^*$ versus the weighting factor $\mathcal{I}^*$ for the parameters in Table~\ref{tab:params}. The driving mode transitions continuously from ACC-dominant to CC-dominant as $\mathcal{I}^*$ increases, reducing $\theta^*$ and raising throughput; the braking-based safety constraint caps the weighting factor at $\mathcal{I}^*_{\max}$, beyond which operation is safety-infeasible (shaded).}
  \label{fig:theta_vs_I}
\end{figure}

Figure~\ref{fig:theta_vs_I} shows the equilibrium time headway $\theta^*$ as a function of the weighting factor $\mathcal{I}^*$. As $\mathcal{I}^*$ increases from the ACC-dominant regime ($\mathcal{I}^*\to0$) toward the CC-dominant regime, $\theta^*$ decreases monotonically with the pure-ACC value $\theta^*_{ACC}=\tau+(\eta+l)/v^*$, so that the equilibrium throughput $q^*=1/\theta^*$ increases, consistent with the analysis in Section~\ref{section:equilibrium}. The braking-based safety constraint, however, bounds the weighting factor by $\mathcal{I}^*_{\max}$, as illustrated in Remark~\ref{remark:safety}. Therefore, the operating point is safe for $\mathcal{I}^*\le \mathcal{I}^*_{\max}$ (solid curve in Figure~\ref{fig:theta_vs_I}), whereas the shaded region $\mathcal{I}^*>\mathcal{I}^*_{\max}$ is safety-infeasible (dashed curve in Figure~\ref{fig:theta_vs_I}). The maximum achievable throughput is therefore reached at $\mathcal{I}^*_{\max}$.

\subsection{Comparison of the CC, ACC, and Unified Models}\label{section:sim_platoon}

This subsection compares the platoon behavior under three operating modes, i.e., the pure CC mode ($\mathcal{I}\to 1$), the pure ACC mode ($\mathcal{I}\to 0$), and the proposed unified model ($0<\mathcal{I}<1$). The goal is to illustrate the distinct characteristics of each mode and to verify that the unified model with the proposed switching design improves throughput and string stability over the pure ACC mode while remaining collision-free.

\begin{figure}
  \centering
  \includegraphics[width=\columnwidth]{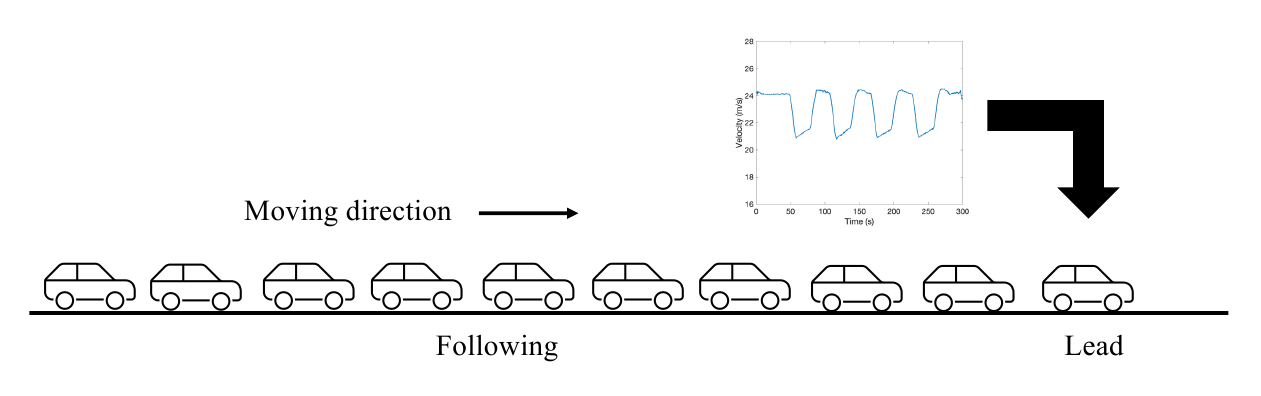}
  \vskip-10pt
  \caption{Schematic of the simulated platoon of $N=10$ vehicles. The lead vehicle follows the field-measured oscillatory speed profile, and the nine following vehicles are governed by the car-following dynamics under comparison.}
  \label{fig:platoon_schematic}
\end{figure}

As illustrated in Figure~\ref{fig:platoon_schematic}, the platoon consists of $N=10$ vehicles, in which the lead vehicle follows a field-measured, oscillatory speed profile collected in~\citep{gunter2020are}. The unified car-following dynamics in~\eqref{eq:switch} are integrated using the forward Euler scheme with time step $\Delta t=0.1$~s over a horizon of $300$~s. For the unified mode, the switching parameters follow the throughput-priority design of Section~\ref{section:optimal}, i.e., $\tau_c=\tau_c^{opt}=t_r$, equivalently $\mathcal{I}^*=\mathcal{I}^*_{\max}$. 

\begin{figure}
  \centering
  \begin{subfigure}{0.33\textwidth}
    \includegraphics[width=\textwidth]{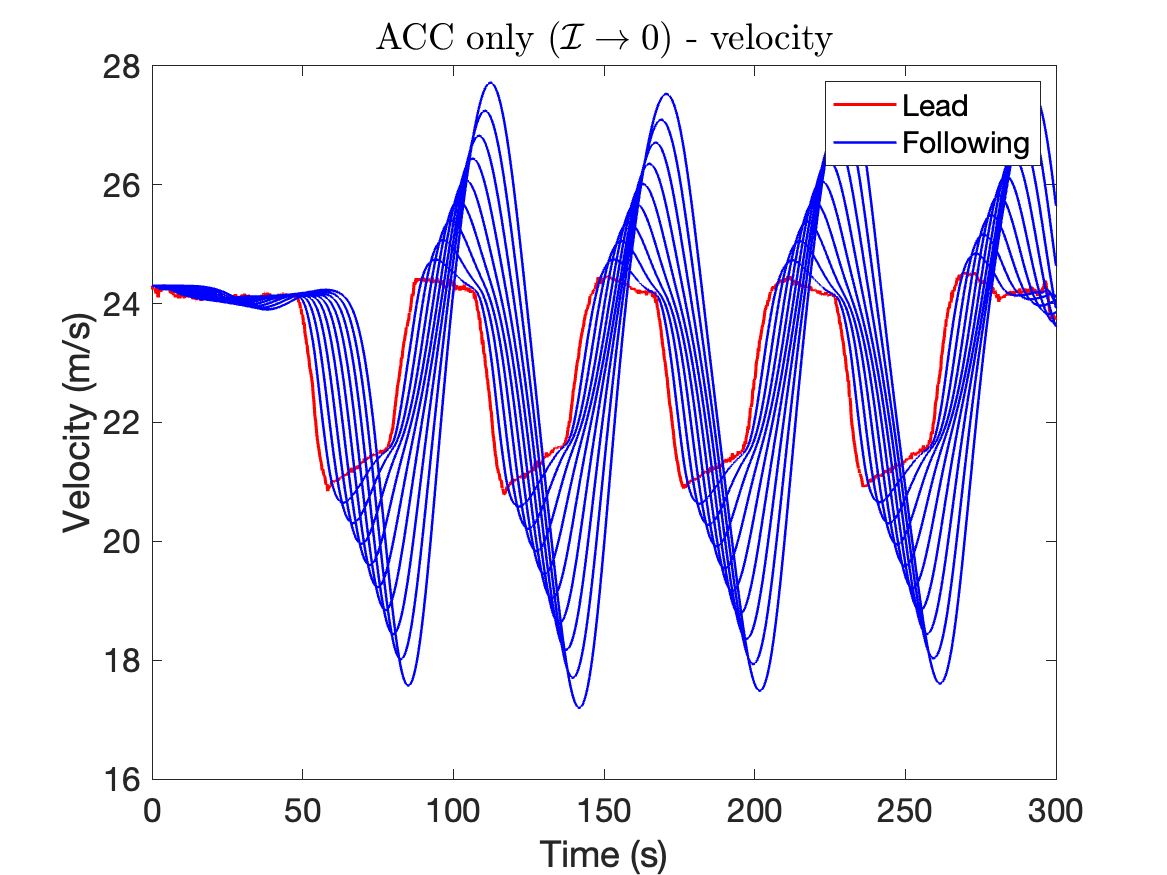}
    \caption{ACC only ($\mathcal{I}\to 0$)}
  \end{subfigure}\hfill
  \begin{subfigure}{0.33\textwidth}
    \includegraphics[width=\textwidth]{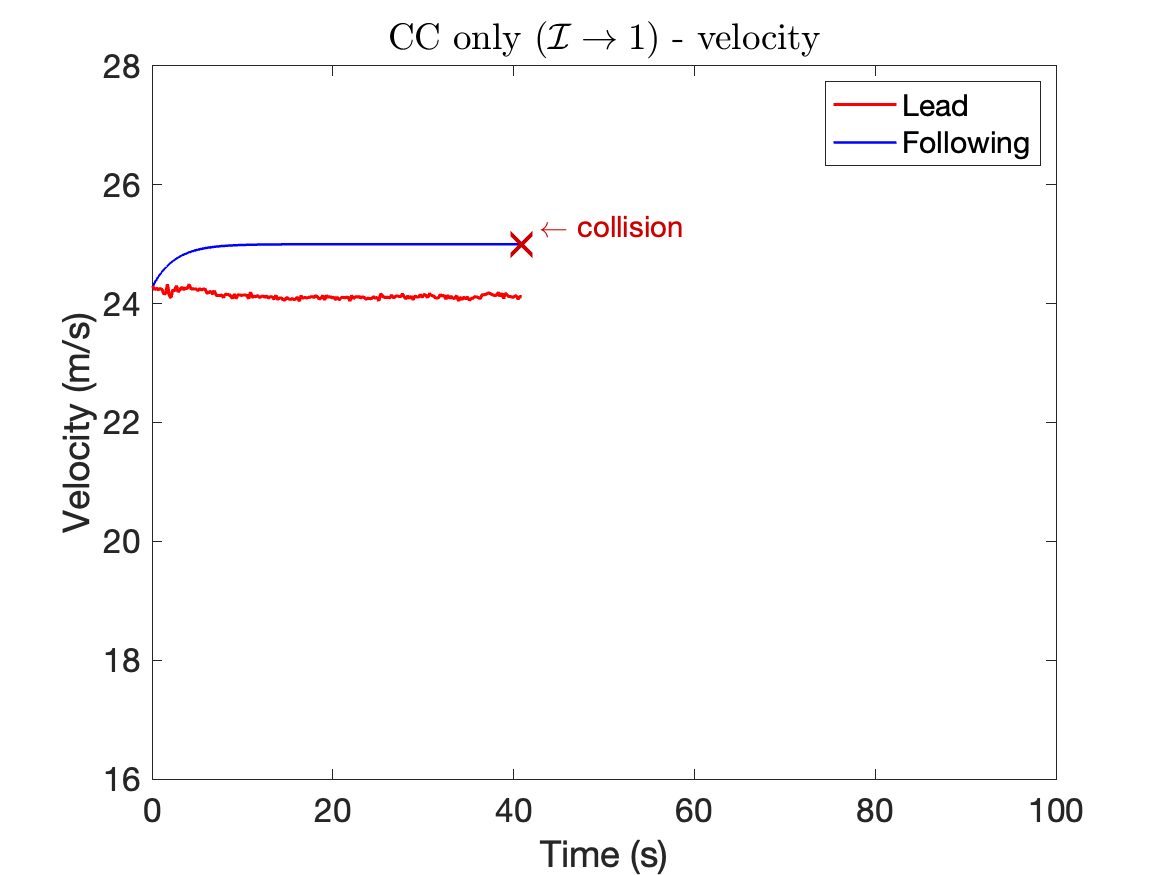}
    \caption{CC only ($\mathcal{I}\to 1$)}
  \end{subfigure}\hfill
  \begin{subfigure}{0.33\textwidth}
    \includegraphics[width=\textwidth]{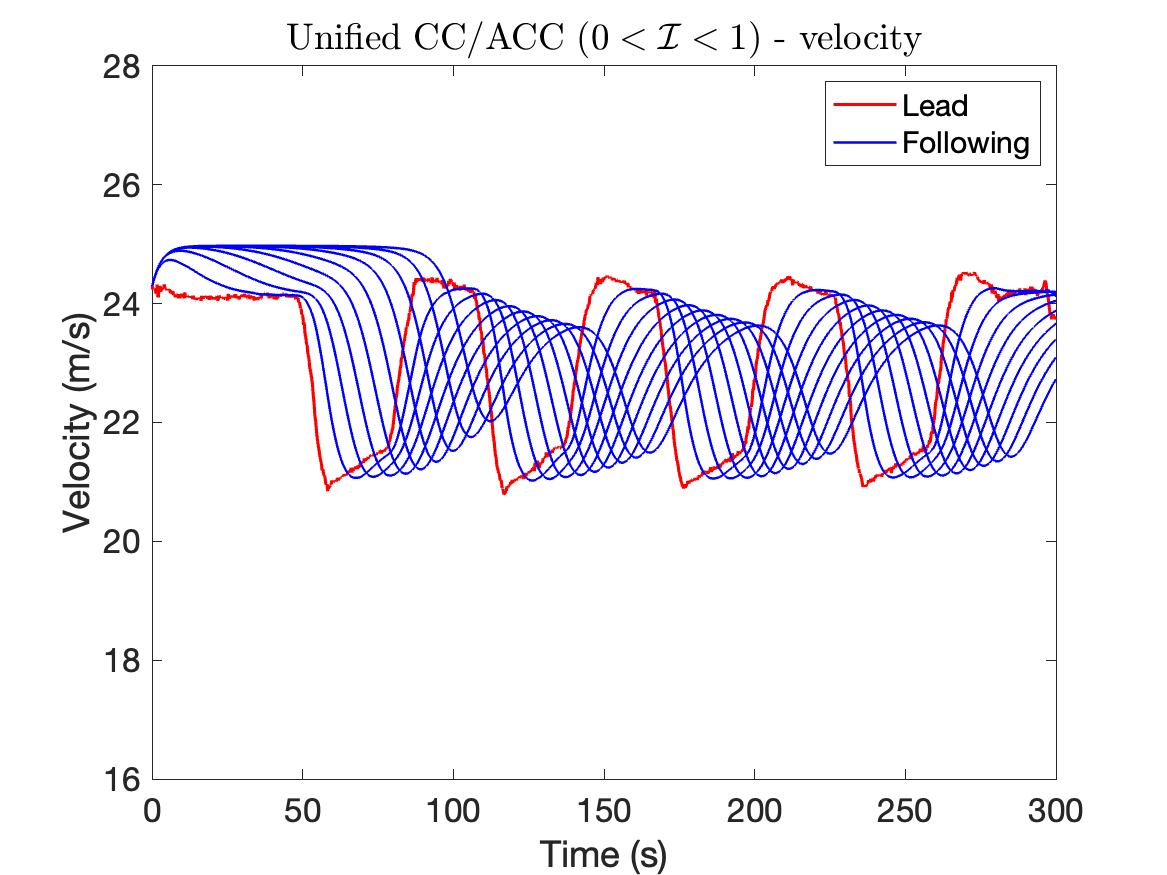}
    \caption{Unified ($0<\mathcal{I}<1$)}
  \end{subfigure}
  \caption{Platoon velocity under the three operating modes.}
  \label{fig:cmp_velocity}
\end{figure}

Figure~\ref{fig:cmp_velocity} compares the velocity profiles. Under the pure ACC mode, the lead-vehicle disturbance is amplified as it propagates upstream along the platoon: the lead vehicle oscillates between about $21$ and $24.5~\mathrm{m/s}$, whereas the speed reaches about $17$--$28~\mathrm{m/s}$ at the end of the platoon, reflecting that the pure-ACC mode is string unstable. Under the pure CC mode, the followers converge to the desired speed $v_d=25.0~\mathrm{m/s}$ and are insensitive to the lead-vehicle disturbance, since the CC mode regulates the speed without measuring the gap and chances of collisions. Under the unified mode, the following vehicles follow the lead vehicle and the oscillations propagate with dissipation, remaining bounded between about $21$ and $25~\mathrm{m/s}$, indicating the unified mode is string stable.

\begin{figure}
  \centering
  \begin{subfigure}{0.33\textwidth}
    \includegraphics[width=\textwidth]{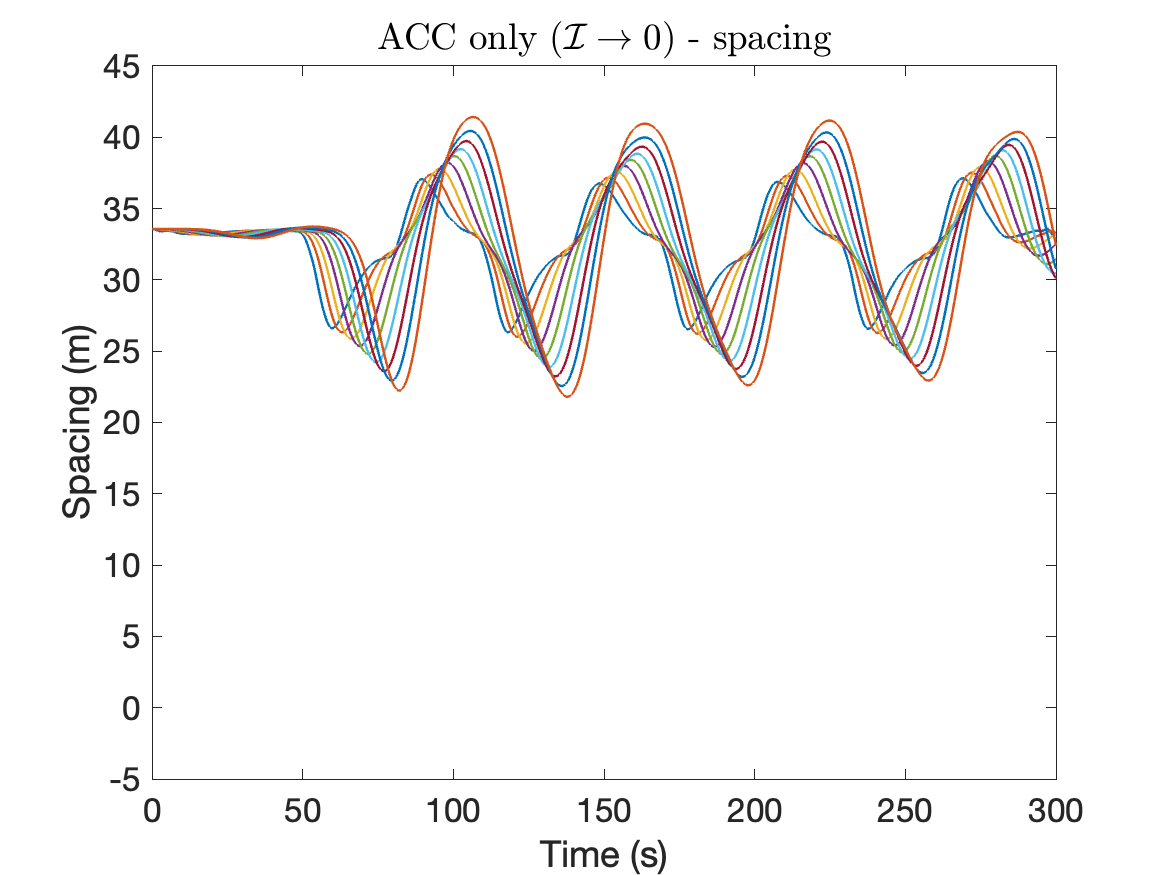}
    \caption{ACC only ($\mathcal{I}\to 0$)}
  \end{subfigure}\hfill
  \begin{subfigure}{0.33\textwidth}
    \includegraphics[width=\textwidth]{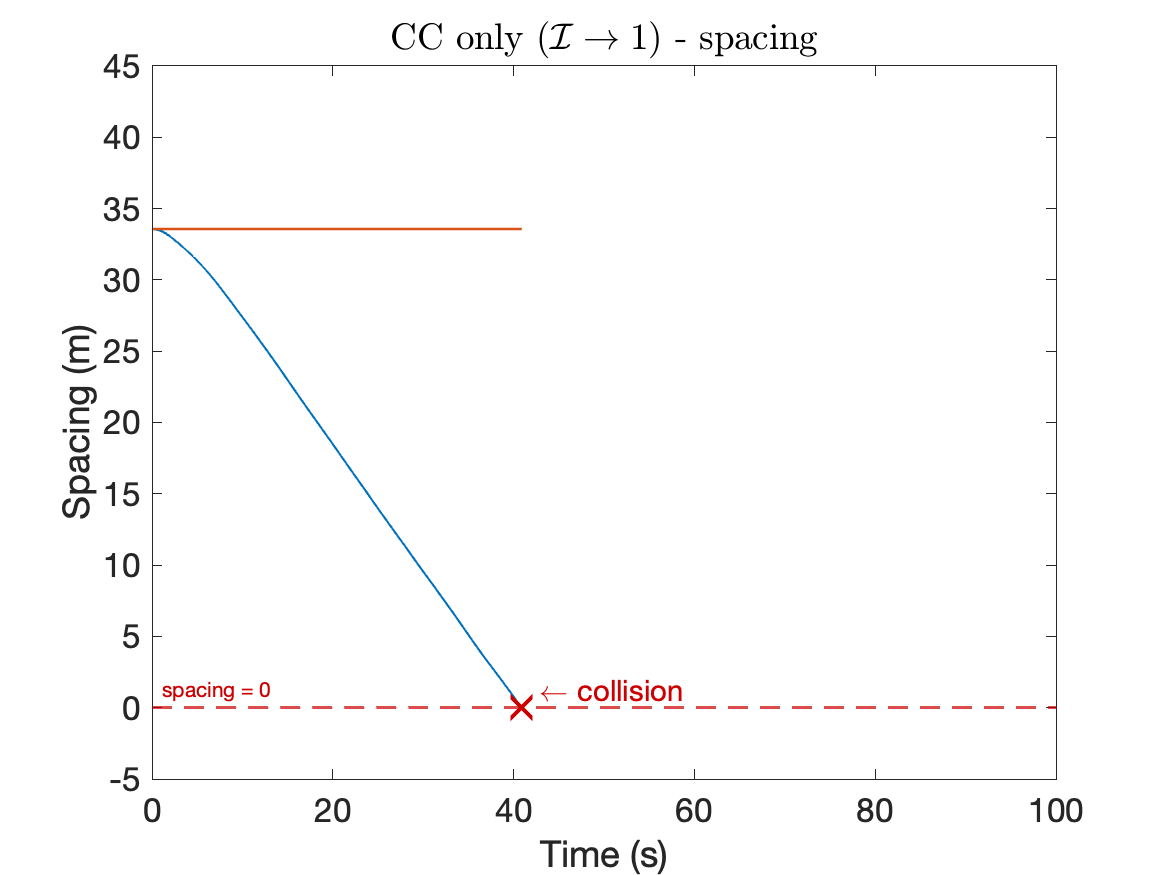}
    \caption{CC only ($\mathcal{I}\to 1$)}
  \end{subfigure}\hfill
  \begin{subfigure}{0.33\textwidth}
    \includegraphics[width=\textwidth]{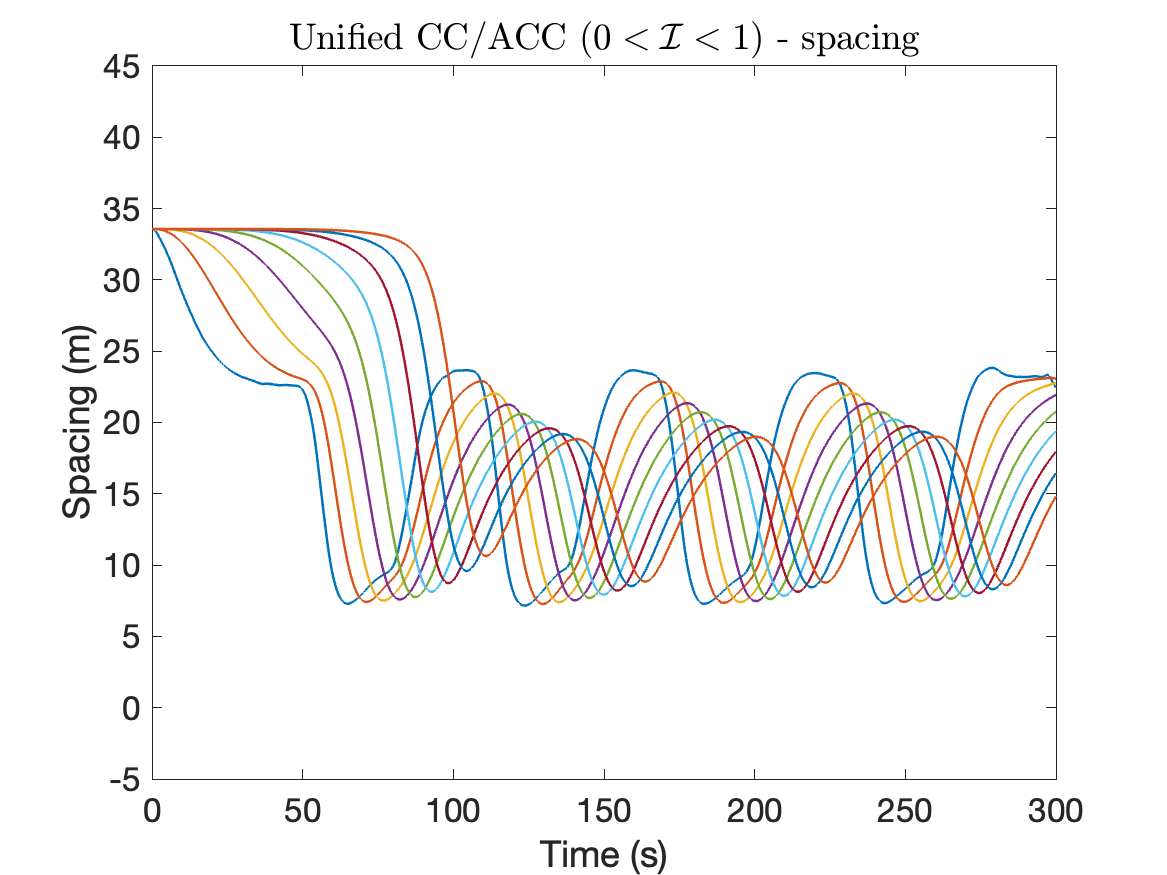}
    \caption{Unified ($0<\mathcal{I}<1$)}
  \end{subfigure}
  \caption{Inter-vehicle spacing under the three operating modes.}
  \label{fig:cmp_spacing}
\end{figure}

Figure~\ref{fig:cmp_spacing} compares the inter-vehicle spacing. The pure ACC platoon maintains large, conservative gaps oscillating around the ACC equilibrium spacing of about $32$~m. In the pure CC mode, the first follower travels at the desired speed, which exceeds the average speed of the lead vehicle, so its spacing gap decreases to zero at $t=40.9$~s, i.e., a collision occurs and the simulation is terminated. Under the unified mode, the spacing gaps are significantly smaller than in the pure-ACC mode, while the spacing remains positive at all times, indicating that the throughput is increased without compromising safety.

\begin{figure}
  \centering
  \begin{subfigure}{0.33\textwidth}
    \includegraphics[width=\textwidth]{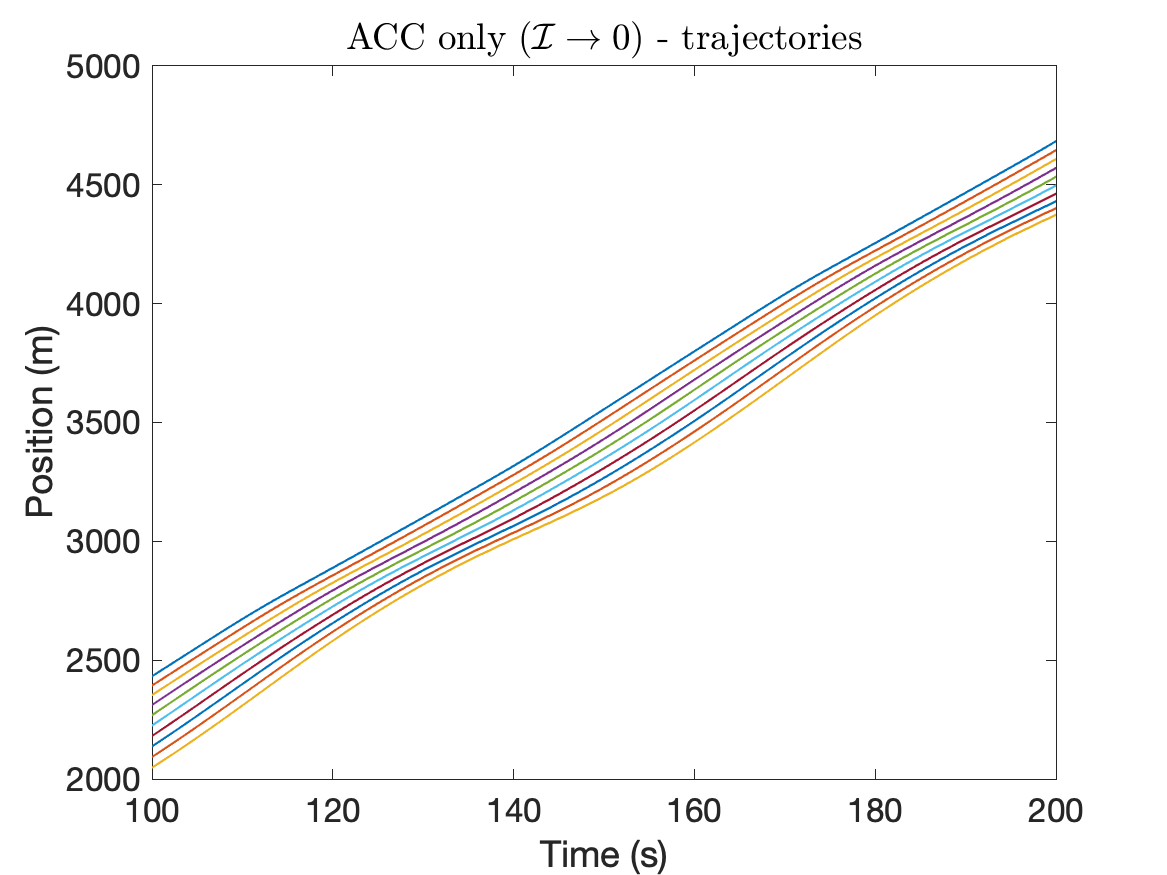}
    \caption{ACC only ($\mathcal{I}\to 0$)}
  \end{subfigure}\hfill
  \begin{subfigure}{0.33\textwidth}
    \includegraphics[width=\textwidth]{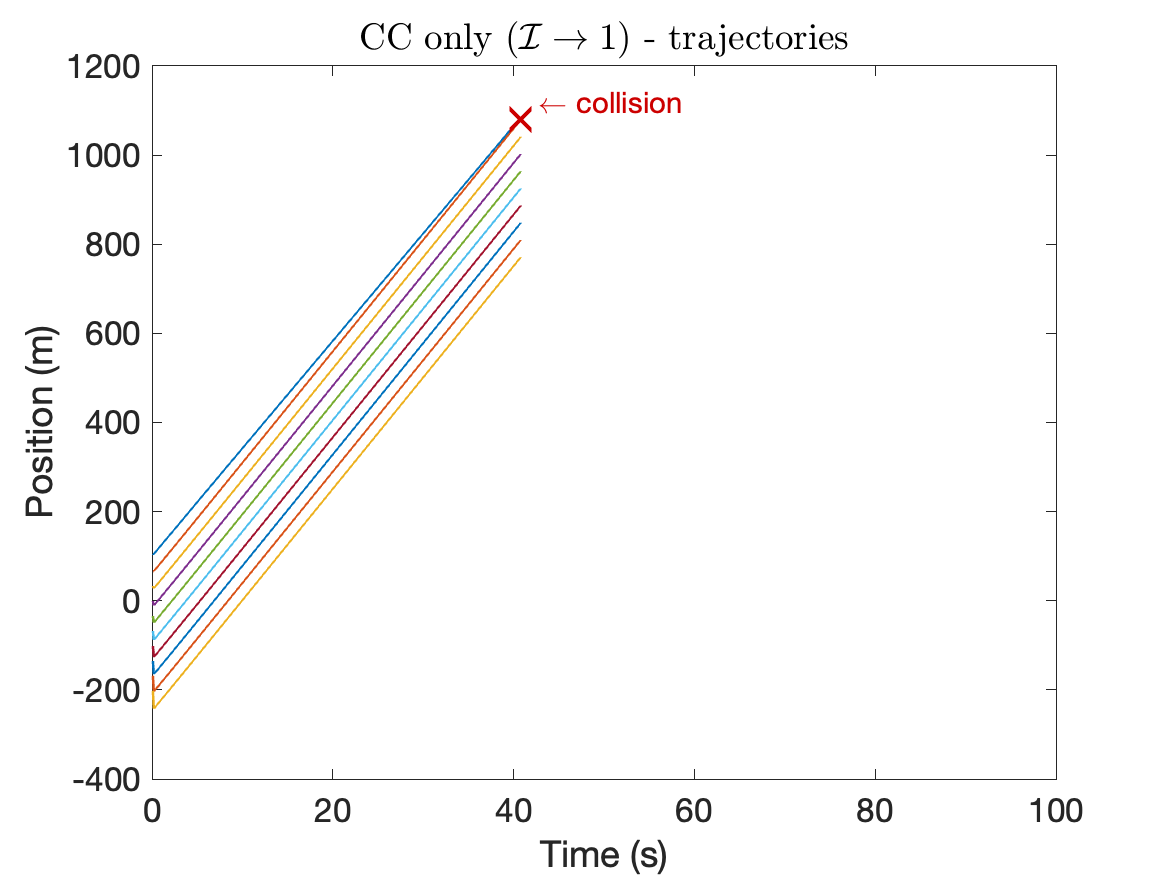}
    \caption{CC only ($\mathcal{I}\to 1$)}
  \end{subfigure}\hfill
  \begin{subfigure}{0.33\textwidth}
    \includegraphics[width=\textwidth]{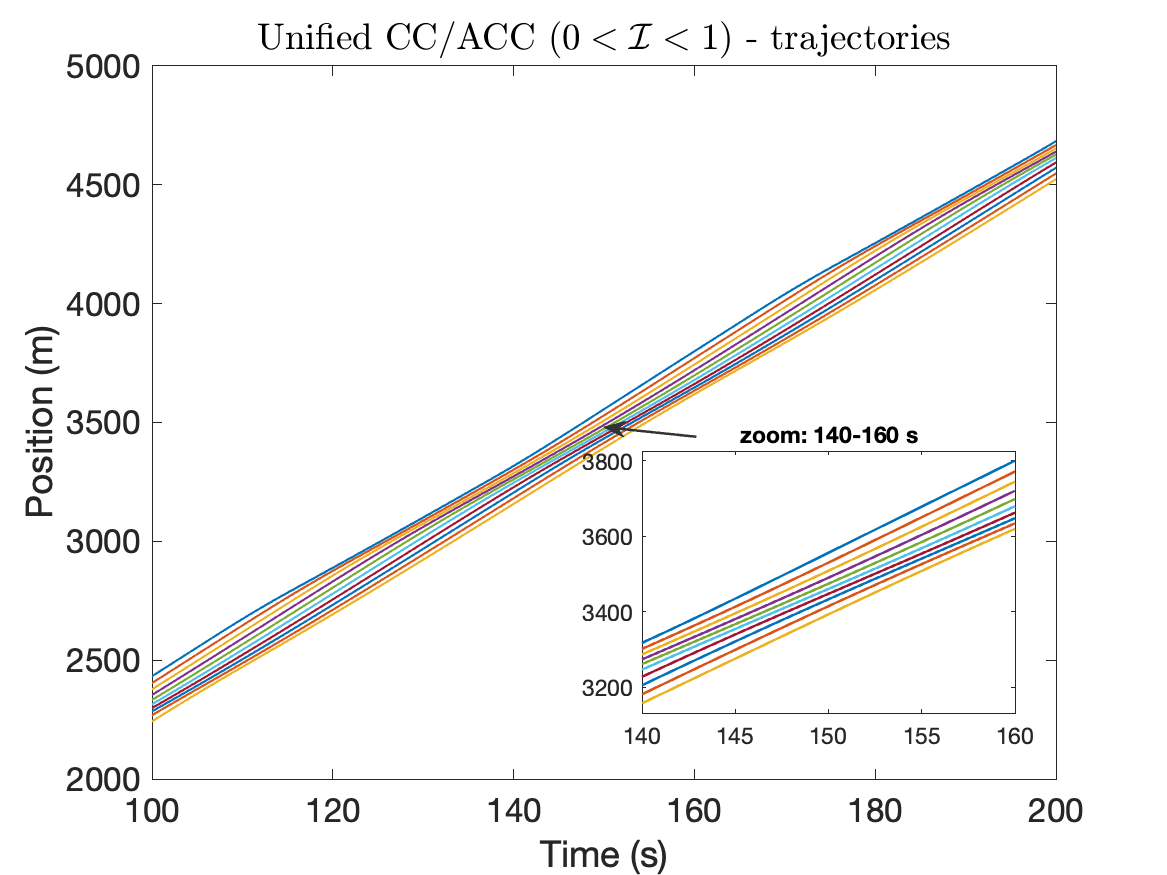}
    \caption{Unified ($0<\mathcal{I}<1$)}
  \end{subfigure}
  \caption{Vehicle trajectories under the three operating modes.}
  \label{fig:cmp_trajectory}
\end{figure}

Figure~\ref{fig:cmp_trajectory} shows the vehicle trajectories over a $100$~s window. The pure-ACC platoon occupies the longest road segment, as its large inter-vehicle gaps stretch the platoon and the stop-and-go oscillations remain visible along the trajectories. In the pure CC mode, the trajectories end up with collisions, and the simulation is terminated thereafter. The unified platoon exhibits denser trajectories, showing a more compact formation than in the pure-ACC scenario and thus enabling a higher throughput, consistent with Figure~\ref{fig:theta_vs_I}.

\begin{figure}
  \centering
  \includegraphics[width=0.5\columnwidth]{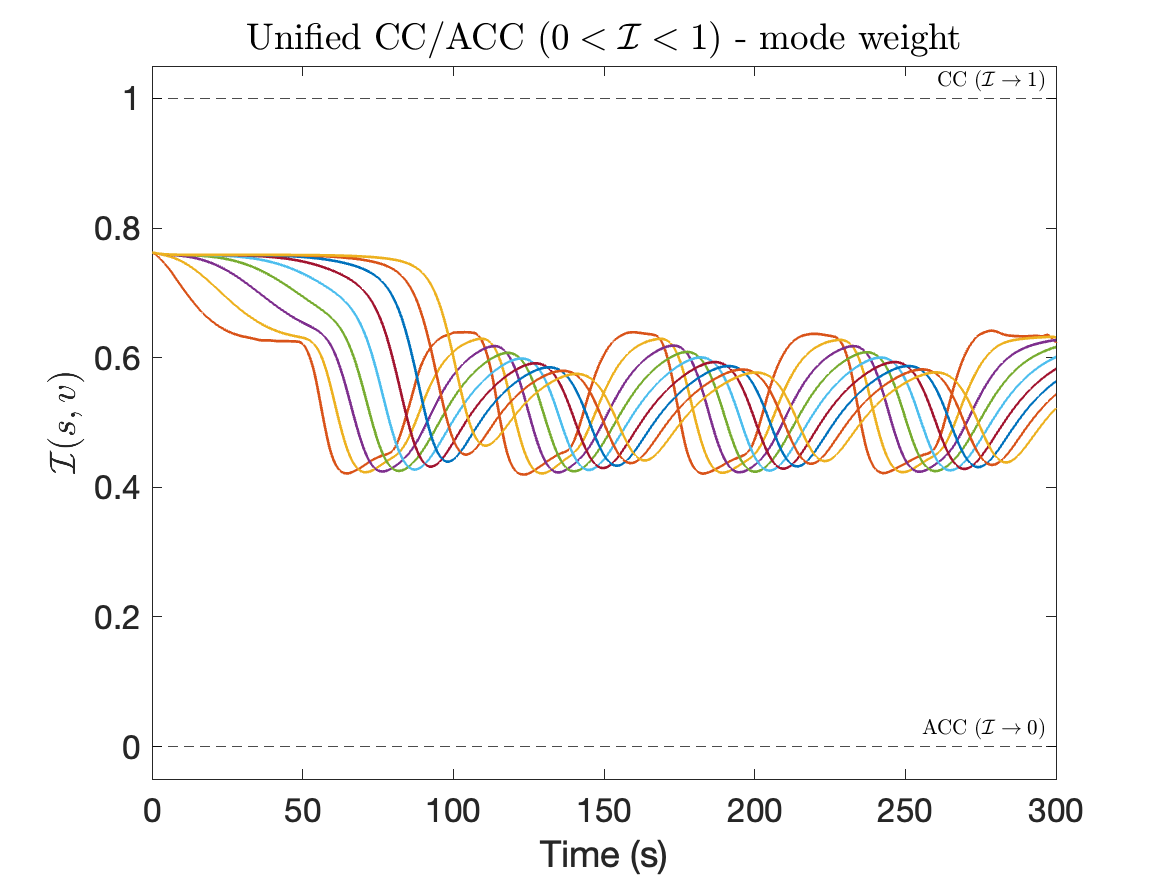}
  \caption{Weighting factor $\mathcal{I}(s,v)$ over time for each follower under the unified model.}
  \label{fig:Iweight}
\end{figure}

Figure~\ref{fig:Iweight} shows the weighting factor $\mathcal{I}(s,v)$ of each follower under the unified mode. In the early stage, the spacing gap is large and stays above the switching threshold, so the weighting factor starts CC-dominant at $\mathcal{I}\approx0.76$. As the spacing approaches its equilibrium value, the weighting factor converges to bounded oscillations around $\mathcal{I}\approx0.53$, indicating a similar utilization of the CC and ACC modes.

\subsection{Comparisons of the Switching Threshold}\label{section:sim_tauc}

To examine how the switching threshold influences the balance between the CC and ACC modes, we compare the unified model under three representative choices of $\tau_c$:
\begin{enumerate}
\item \emph{Throughput-priority} switching threshold ($\tau_c^{opt}=t_r=0.50$~s): defined as the smallest threshold admitted by the braking-based minimum-gap constraint of Section~\ref{section:optimal}. As the most CC-dominant feasible operating point, it yields the highest throughput while still guaranteeing safety.
\item \emph{Safety-priority} switching threshold ($\tau_{c,\mathrm{stab}}=1.72$~s): defined as the largest threshold that still keeps the platoon string stable. The value $\tau_{c,\mathrm{stab}}=1.72$~s is calculated with the method as indicated in Remark~\ref{remark:tauc_stab}. Therefore, any larger threshold results in the platoon string unstable.
\item \emph{Commercially available} switching threshold ($\tau_{c,\mathrm{comm}}=4.25$~s): a value representative of commercially available ACC vehicles, which cruise at a set speed until a preceding vehicle is acquired within the sensor range and then regulate the gap~\citep{milanes2014modeling}. Long-range radars used by such systems detect preceding vehicles up to about $200$~m ahead~\citep{mohammed2024comparative}. Within this envelope, we adopt a representative switching spacing of $s_c=100$~m, giving the equivalent threshold $\tau_{c,\mathrm{comm}}=(s_c-s_0)/v^*=4.25$~s. This strongly ACC-dominant setting lies well inside the string-unstable regime.
\end{enumerate}

To quantify and evaluate the impacts of the switching threshold, we use two commonly-used evaluation metrics: the average speed variation (ASV) adopted from~\citep{wang2023general} to quantify the traffic smoothness of the platoon, and the throughput to measure its traffic efficiency. Specifically, the ASV is given by
\begin{align}
\mathrm{ASV}=\frac{1}{M T}\sum_{i=1}^{M}\sum_{t=1}^{T}\left|v_i^t-v^*\right|,
\end{align}
where $M=9$ is the number of followers, $T$ the number of time steps, $v_i^t$ the speed of follower $i$ at step $t$, and $v^*$ the equilibrium speed; a larger value indicates greater oscillations. Traffic efficiency is measured by the throughput
\begin{align}
q=\frac{3600}{\bar h},\qquad
\bar h=\frac{1}{M\,T}\sum_{i=1}^{M}\sum_{t=1}^{T}\frac{s_i^t+l}{v_i^t},
\end{align}
where $s_i^t$ is the bumper-to-bumper spacing, $l$ the vehicle length, $(s_i^t+l)/v_i^t$ the time headway, $\bar h$ its mean over the $M$ followers and $T$ time steps, and the factor $3600$ converts veh/s to veh/h. 

\begin{figure}
  \centering
  \begin{subfigure}{0.33\textwidth}
    \includegraphics[width=\textwidth]{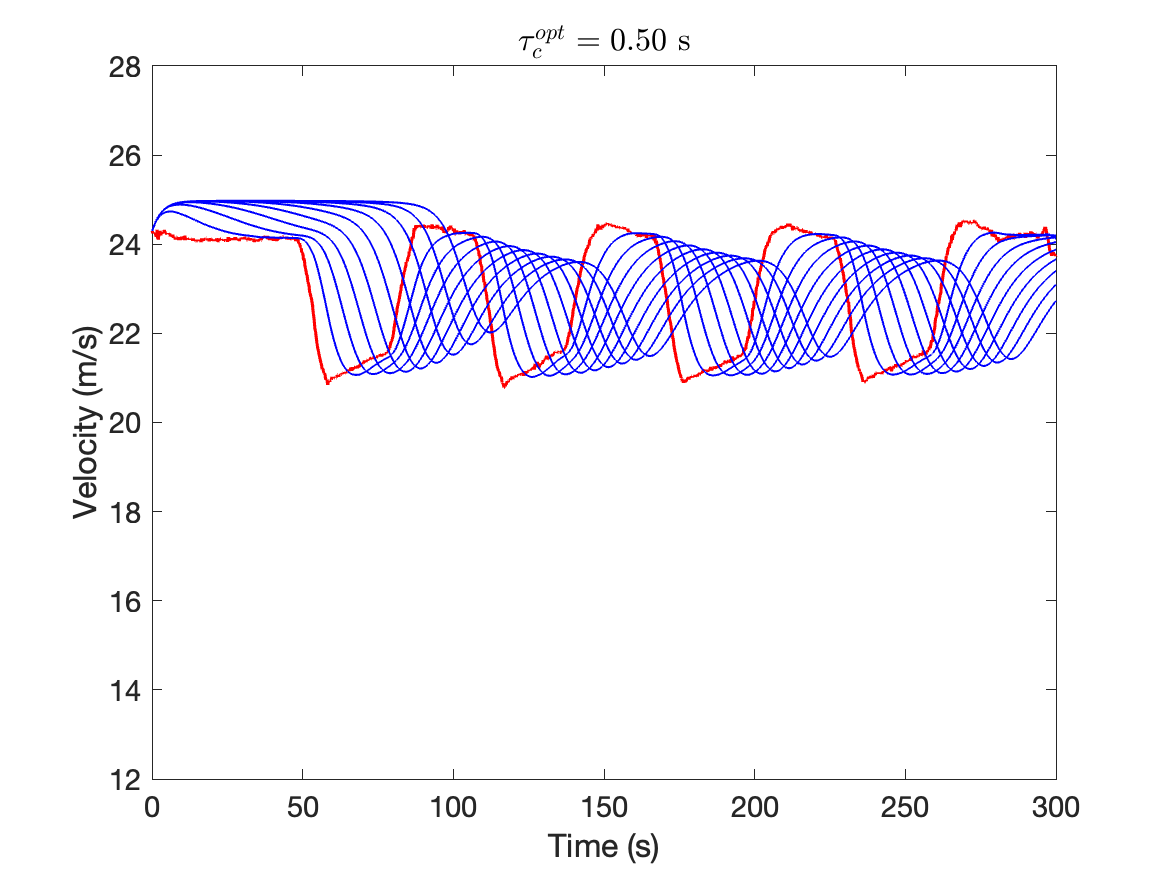}
    \caption{Throughput-priority: $\tau_c^{opt} = 0.50 s$}
  \end{subfigure}\hfill
  \begin{subfigure}{0.33\textwidth}
    \includegraphics[width=\textwidth]{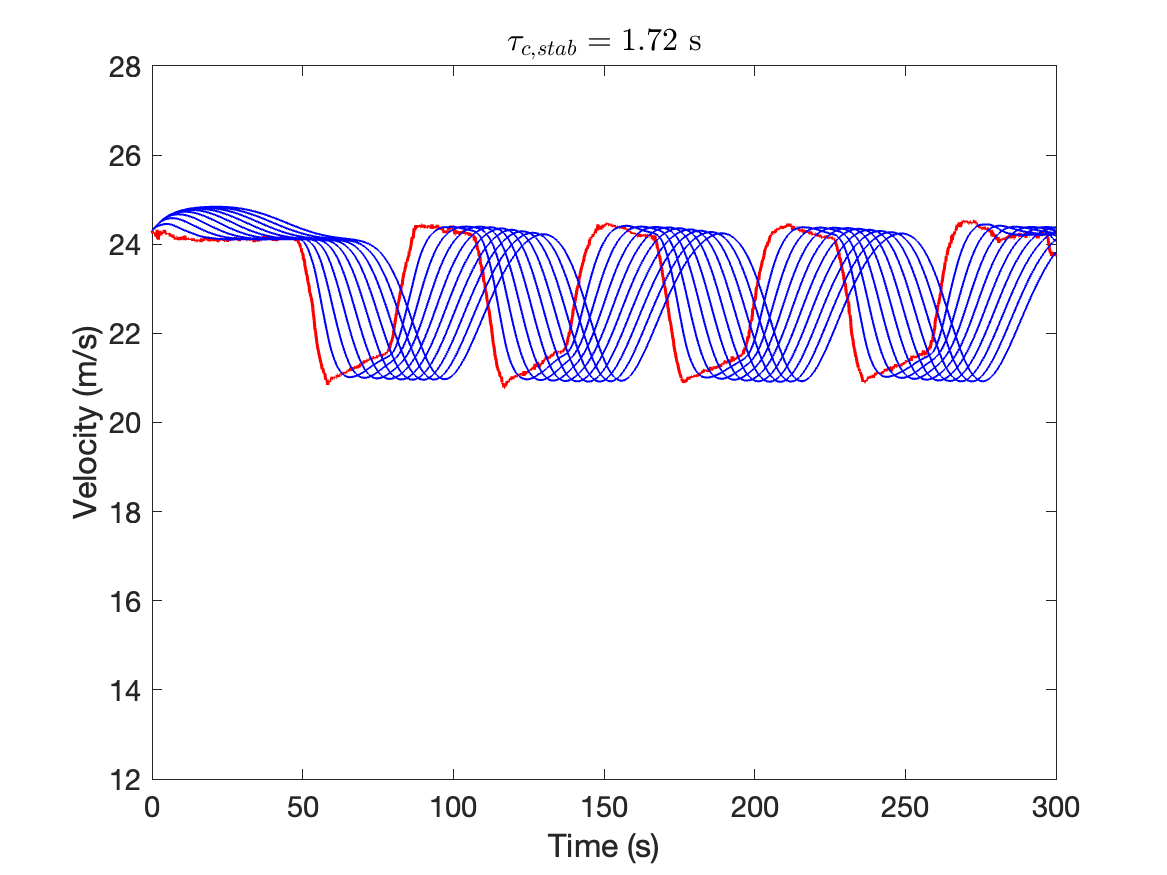}
    \caption{Safety-priority: $\tau_{c,\mathrm{stab}} = 1.72 s$}
  \end{subfigure}\hfill
  \begin{subfigure}{0.33\textwidth}
    \includegraphics[width=\textwidth]{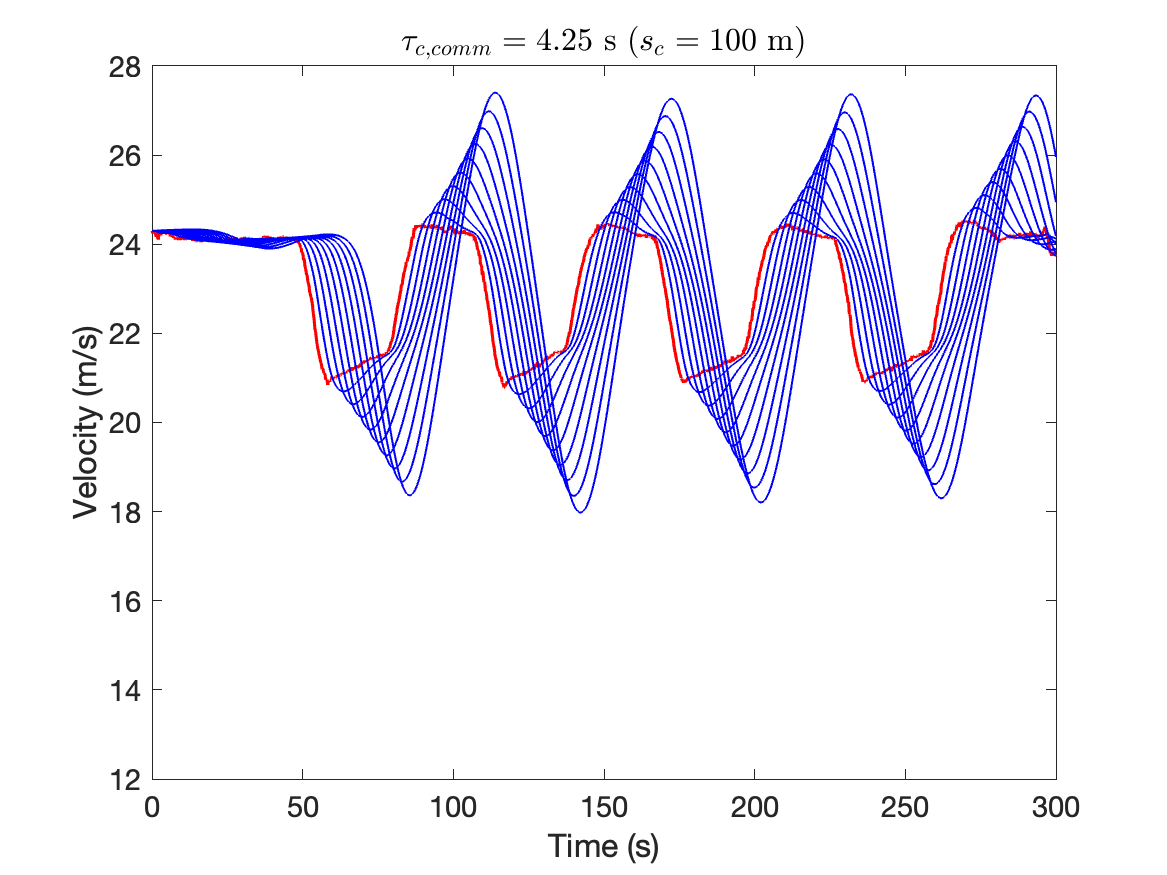}
    \caption{Commercial: $\tau_{c,\mathrm{comm}}=4.25 s$ }
  \end{subfigure}
  \caption{Follower speed profiles under the three switching thresholds. The throughput-priority and safety-priority thresholds attenuate the propagation of speed disturbances along the platoon, whereas the commercially available switching threshold amplifies disturbance propagation.}
  \label{fig:tauc_velocity}
\end{figure}

Figure~\ref{fig:tauc_velocity} shows the follower speed profiles under the three thresholds, and Table~\ref{tab:tauc} summarizes the corresponding ASV and throughput for a simulated 10-vehicle platoon, together with their relative change with respect to the commercially available switching threshold baseline ($\tau_{c,\mathrm{comm}}$), i.e., $\Delta\mathrm{ASV}=(\mathrm{ASV}-\mathrm{ASV}_{\mathrm{comm}})/\mathrm{ASV}_{\mathrm{comm}}\times100\%$ and $\Delta q=(q-q_{\mathrm{comm}})/q_{\mathrm{comm}}\times100\%$.

Under the commercially available switching threshold, the perturbation is amplified along the platoon. Its ASV is larger than that under the throughput-priority and safety-priority thresholds, indicating that the commercially available switching threshold leads to more pronounced string instability. Its throughput is also the lowest compared with the other two switching thresholds, meaning that commercially available ACC vehicles may negatively impact traffic flow~\citep{shang2021impacts}, likely due to their large switching threshold. In contrast, both $\tau_c^{opt}$ and $\tau_{c,\mathrm{stab}}$ dissipate the perturbation by reducing the ASV by 39.7\% and 34.7\%, respectively, and increase the throughput by 58.6\% and 16.4\%, respectively.

\subsection{Sensitivity Analysis}

The unified model is also influenced by the transition steepness factor $\alpha$, as indicated in Remark~\ref{remark:alpha}. Next, we conduct a sensitivity analysis to understand how the threshold $\tau_c$ and the steepness $\alpha$ jointly affect the traffic flow characteristics of the platoon. We simulate the 10-vehicle platoon on a grid of $(\tau_c,\alpha)$ values, where $\tau_c \in [0.5, 4.5]$ and $\alpha \in [0.01, 0.1]$, and record the ASV and the throughput at each point.

\begin{figure}
  \centering
  \includegraphics[width=\textwidth]{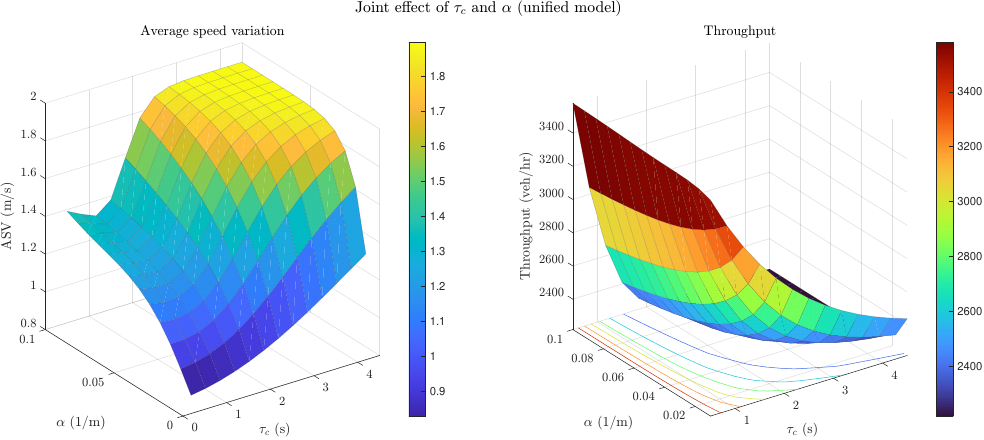} 
  \caption{Joint effect of the switching threshold $\tau_c$ and sigmoid steepness $\alpha$ on average speed variation (ASV, left) and throughput (right). Reducing the switching threshold to $\tau_c=t_r$ and decreasing the sigmoid steepness $\alpha$ improve traffic performance by reducing speed fluctuations and increasing throughput.}
\label{fig:tauc_alpha}
\end{figure}

Figure~\ref{fig:tauc_alpha} shows the ASV and throughput results on 3-D surfaces. The ASV surface (left) shows a lower ASV when $\tau_c$ and $\alpha$ are both small, indicating weaker stop-and-go oscillations in the traffic. Increasing either parameter raises the ASV and drives the platoon toward instability. The result confirms Remark~\ref{remark:alpha} that string stability is influenced by both $\tau_c$ and $\alpha$. The throughput surface (right) shows that increasing $\tau_c$ decreases the throughput. Therefore, the two surfaces suggest that a small $\tau_c$ down to $t_r$ together with a small $\alpha$ results in both the lowest ASV and the highest throughput, so that the platoon is string stable and efficient simultaneously.

\begin{table}
\centering
\caption{ASV and throughput $q$ for a simulated 10-vehicle platoon under three switching thresholds, with the relative change against the commercially available threshold baseline ($\tau_{c,\mathrm{comm}}$).}
\label{tab:tauc}
\begin{tabular}{lccccc}
\toprule
Threshold & $\tau_c$ [s] & ASV [m/s] & $\Delta$ASV & $q$ [veh/h] & $\Delta q$ \\
\midrule
$\tau_c^{opt}$          & $0.50$ & $1.086$ & $-39.7\%$ & $3552$ & $+58.6\%$ \\
$\tau_{c,\mathrm{stab}}$ & $1.72$ & $1.177$ & $-34.7\%$ & $2608$ & $+16.4\%$ \\
$\tau_{c,\mathrm{comm}}$  & $4.25$ & $1.802$ & ---       & $2240$ & ---       \\
\bottomrule
\end{tabular}
\end{table}

\section{Conclusion}\label{section5}

ACC vehicles represent one of the earliest and most widely deployed forms of automated driving technology. However, field experiments have shown that commercially available ACC systems may degrade traffic flow by reducing string stability and throughput. Existing studies have attempted to mitigate these effects by modifying ACC control algorithms or introducing additional control inputs. Nevertheless, most existing car-following models describe ACC dynamics alone and neglect the coexistence and transition between CC and ACC modes. To address this gap, we propose a unified dynamical model that continuously interpolates between CC and ACC modes through a sigmoid weighting function. Instead of modifying the ACC controller itself, the proposed approach improves traffic performance by designing the mode-switching mechanism. Equilibrium and string stability analyses formulate the switching design as bounds on the weighting factor, enabling the safe and string-stable operating region and optimal switching thresholds to be derived under throughput-priority and safety-priority criteria. We further characterize the effects of the switching threshold and sigmoid steepness on string stability.

The results demonstrate that, compared with pure ACC operation, the proposed unified model with throughput-priority design improves both throughput and string stability while maintaining collision-free operation, whereas pure CC operation achieves high efficiency at the expense of safety. Comparisons among throughput-priority, safety-priority, and commercially implemented switching thresholds reveal a fundamental tradeoff between efficiency and safety margin. A smaller switching threshold provides higher throughput and stronger disturbance attenuation, while a larger threshold provides a more conservative operating regime. The throughput-priority design reduces ASV by 39.7\% and increases throughput by 58.6\%, whereas the safety-priority design reduces ASV by 34.7\% and increases throughput by 16.4\%. Sensitivity analysis further shows that string stability requires both the switching threshold and sigmoid steepness to remain sufficiently small, while throughput is primarily governed by the switching threshold. These findings suggest that excessively large switching thresholds in current ACC systems may contribute to their adverse impacts on traffic flow and that improved traffic performance can be achieved by shifting the switching mechanism toward a safer and more string-stable regime.

Future work will extend the unified model from homogeneous vehicle platoons to heterogeneous and mixed-autonomy traffic environments. Experimental validation using field data and large-scale network simulations will also be conducted to evaluate the practical feasibility and system-level impacts of the proposed switching strategy.

\bibliographystyle{unsrtnat}
\bibliography{mybibfile}

\end{document}